\documentclass{aa}  

\usepackage{amsmath}
\usepackage{natbib}
\usepackage{graphicx}
\graphicspath{{1_Trs_FMW_AW_ideal_Kx0_vars}, {2_Trs_FMW_AW_ideal_Ky_vars}, {3_Trs_FMW_AW_ideal_R_vars}, {4_Trs_FMW_AW_diffusion_eta_vars}, {5_Application_to_threads}}
\usepackage{txfonts}
\usepackage{lipsum}
\usepackage{subcaption} 
\usepackage{lscape}
\usepackage{placeins}
\usepackage{hyperref}
\usepackage{bm}
\usepackage{lineno}
\nolinenumbers

\begin{document}

   \title{Wave coupling in partially ionized plasmas with shear flows}

   \subtitle{I. Fast-to-Alfv\'en transformation}

   \author{Miquel Cantallops\inst{1,2}
        \and Roberto Soler\inst{1,2}\fnmsep}

   \institute{Departament de Física, Universitat de les Illes Balears, 07122 Palma de Mallorca, Spain\\
            \email{miquel.cantallops@uib.cat}
            \and Institut d’Aplicacions Computacionals de Codi Comunitari (IAC3), Universitat de les Illes Balears, 07122 Palma de Mallorca,
            Spain\\ }

   \date{Received September 30, 20XX}
 
  \abstract{We theoretically investigate the interplay between magnetohydrodynamic (MHD) waves and shear flows in a partially ionized solar plasma, focusing on the energy exchange mediated by the flow and the transformation between wave modes. We consider a  simple model composed of a uniform partially ionized plasma with a straight magnetic field. A shear flow is present in the direction of the magnetic field with a velocity that varies linearly across the magnetic field. The linearized MHD equations in the single-fluid approximation are used, which include the ambipolar diffusion term due to ion-neutral collisions. A nonmodal approach is adopted, in order to convert the flow spatial inhomogeneity into a temporal one, adding a temporal dependence into the component of the wavevector in the direction of the flow inhomogeneity. A system of three ordinary differential equations is derived, which generally governs the temporal evolution of the coupled MHD waves, their interaction with the shear flow, and their ambipolar damping. Numerical solutions are computed to study the coupling and mutual transformation between the fast magnetosonic wave and the Alfv\'en wave. A detailed parameter study is conducted, demonstrating how  the energy transfer and mode transformation are affected. The role of ambipolar diffusion is investigated by comparing the results of the ideal case with those obtained when ambipolar diffusion is included. It is found that ambipolar diffusion can significantly affect the efficiency of the energy exchange between modes and introduces a new coupling mechanism. Additionally, a specific application to solar prominence threads is included, showing that wave coupling and energy exchange can occur within these  and other similar structures in the solar atmosphere.}


   \maketitle

\section{Introduction}

 Magnetohydrodynamic (MHD) waves are heavily affected when propagating in a medium with nonuniform properties, such as the solar atmosphere. One direct consequence of nonuniformity is the coupling between the various MHD wave modes \citep[see, e.g.,][]{callybogdan2024}. Owing to the coupling,  the  perturbations associated with  MHD waves generally display mixed properties of the pure modes depending on position and/or time  \citep[see, e.g.,][]{solergoossens2024}. However, it has been shown  that the distinct pure modes remain definable \citep[see, e.g.,][]{raboonik2024}. As the plasmas in the lower atmosphere and solar prominences are in a state of partial ionization, effects caused by ion-neutral collisions, like ambipolar diffusion,  can also affect wave propagation and coupling in such plasmas  \citep{2024SolerMHDWavesinPIP}.

In the vertically stratified lower solar atmosphere, the profiles of sound and Alfv\'en speeds  converge at the so-called equipartition layer. This convergence facilitates coupling between the distinct magnetosonic modes. In addition, in the presence of inclined magnetic fields, conversion to Alfv\'en waves is also possible. Theoretical studies have extensively explored this process, which may be of particular importance above sunspots \citep[see, e.g.,][]{cally2001,Khomenko2015,callybogdan2024}.  In this context, \citet{cally2018} examined the conversion from fast to Alfv\'en modes in the presence of ambipolar diffusion, while \citet{Popescu2021} explored the conversion from fast to slow magnetosonic modes under similar conditions. Notably, their results suggest that the fundamental characteristics of the mode conversion process remain largely unchanged compared to the ideal case. However, the primary difference is the  damping  caused by ambipolar diffusion. A detailed analysis of magnetosonic mode conversion within the two-fluid framework was presented in \citet{cally2023}. They derived a transformation coefficient that includes a correction term owing to ion-neutral collisions, although it was found to be of small importance under solar atmospheric conditions. On the other hand, \citet{cally2019} and \citet{KhomenkoCally2019} studied the fast-to-Alfv\'en mode conversion in an assembly of small-scale flux tubes. They found that the impact of ambipolar diffusion was amplified by the presence of the flux tube structure.

Another nonideal mechanism that can be of potential relevance for wave coupling in the solar plasma is the Hall effect. In the presence of partial ionization, this effect is enhanced by the collisions with neutrals \citep{pandey2008}. Conversion from fast to Alfv\'en modes mediated by the Hall term was studied in \citet{cally2015} and \citet{gonzalez2019}, whereas the coupling between slow and   Alfv\'en modes was analyzed in \citet{Raboonik2019}. While the Hall-mediated conversion from fast to Alfv\'en modes requires significantly high frequencies to be efficient, the conversion from slow to Alfv\'en modes can be relevant at much lower frequencies. The results of these studies are applicable in the photosphere and lower chromosphere.

Resonant absorption can be considered as another type of mode conversion process \citep[see, e.g.,][]{goossens2011}. It describes how MHD wave energy is absorbed into the continuous Alfv\'en or slow spectra. This process depends on gradients across the magnetic field direction. The resonant absorption of Alfv\'enic waves in partially ionized magnetic flux tubes, such as prominence threads or chromospheric spicules, was studied by \citet{soler2009res} using the single-fluid model and by  \citet{Soler2012} in the two-fluid framework. The latter study analyzed the decoupling of ions and neutrals within the extremely thin collisional layer around the resonance. This  decoupling causes dissipation and large velocity gradients that may lead to substantial plasma heating within the narrow layer.

Observations often show that MHD waves and flows are  simultaneously present in the solar atmosphere \citep[e.g.,][]{2007OkamotoCoronal,ofman2008,DePontieu2010,tian2012,ofman2020}. This gives rise to another potential wave coupling mechanism driven by the flows. A  shear flow entails a spatial inhomogeneity and it has been shown that this causes linear coupling and mutual transformation between the MHD waves \citep[see, e.g.,][among others]{1996ChagelishviliCoupling&LinearTransformation,1997ChagelishviliMHDWaves,1997RogavaCouplingHydrodinamicalWaves,2004GogoberidzeLinearCoupling}. A velocity shear causes significant change in the very nature of the eigenvalue problem associated with linear waves. The corresponding eigenvalue problem becomes non-self-adjoint, leading to non-orthogonal eigenfunctions with no independent temporal evolution. Consequently the exponential temporal dependence of the asymptotic normal mode analysis  cannot completely describe the evolution of the coupled waves. Another approach is needed to understand the wave conversion and the energy exchange between modes. The so-called nonmodal analysis \citep[see, e.g.,][]{2007SchmidNonmodal,2012CamporealeNonmodalLinearTheory} has been used in previous studies dealing with shear-induced wave transformations in a wide variety of astrophysical and geophysical contexts \citep[see, e.g.,][]{1994ChagelishviliHydrodynamicStability,1996ChagelishvilAmpVortexDistSF,1997ChagelishviliMHDWaves,1997RogavaCouplingHydrodinamicalWaves,1998RogavaShearVortex,1999RogavaWaveCoupling}. Although there are some works that have studied nonmodal effects on MHD waves in the solar atmosphere and the solar wind  \citep[see][]{1998PoedtsSFWaveCoupling,1999PoedtsShearPhenomenaOverview}, no previous work has analyzed the  shear-induced wave coupling in partially ionized solar plasmas and the role of ambipolar diffusion in the process.

We aim to investigate the wave coupling and energy transfer mechanism driven by shear flows in regions of the solar atmosphere where the plasma is partially ionized, in which the role of ambipolar diffusion can be important. In the present initial paper, we include the derivation of the general governing equations and specifically apply them to explore the mutual transformation between the fast magnetosonic wave and the Alfv\'en wave in the pressureless approximation. Later, in a forthcoming continuation, the investigation will be extended by considering the role of gas pressure, and thus the coupling with the slow magnetosonic wave will be studied.

\section{Background and method}

In order to study the essential features of the shear-induced mode coupling, we consider a simple background  composed of a uniform partially ionized plasma with a constant magnetic field. We use a Cartesian coordinate system so that the background magnetic field is along the $z$-direction, namely $\vec{B}_0=B_0\hat{z}$, with $B_0$ constant. The thermal pressure is also uniform. There is a shear flow parallel to the direction of the magnetic field and whose velocity varies in the $x$-direction, i.e., across the magnetic field. We use a linear variation for the background flow velocity as $\vec{V}_0=Ax\,\hat{z}$, with the constant $A$ measuring the strength of the shear. This configuration, although simple, contains the basic ingredients for our study. This is due to fact that, for small perturbations with wavelengths much shorter than the length scale of the flow variation, any arbitrary flow can locally be approximated by a linear flow. Essentially, this is the same setup as that used in \citet{1998PoedtsSFWaveCoupling}, with the difference that  we allow the plasma to be partially ionized. The role of partial ionization is here included by considering the nonideal mechanism of  ambipolar diffusion.

\subsection{Main equations}

We consider the single-fluid MHD equations that include ambipolar diffusion to account for ion-neutral collisions effects in a partially ionized plasma \citep[see, e.g.,][]{Khomenko2014,2018BallesterPartiallyIonizedPlasma}, namely 
\begin{subequations}
    \begin{align}
        \frac{D\rho}{Dt}&=-\rho\nabla\cdot\vec{v},
        \label{eqn:eq1-a}\\
        \rho\frac{D\vec{v}}{Dt}&=-\nabla p+\vec{J}\times\vec{B} \\
        \frac{\partial\vec{B}}{\partial t}&=\nabla\times\left(\vec{v}\times\vec{B}\right)+\nabla\times\left\{\frac{\mu_0\eta_{A}}{|\vec{B}|^2} \left[\left(\vec{J}\times\vec{B}\right)\times\vec{B} \right] \right\},
        \label{eqn:eq1-c}\\
        \frac{Dp}{Dt}&=\frac{\gamma p}{\rho}\frac{D\rho}{Dt}+\left(\gamma-1\right)\mu_0\eta_A\left|\frac{\vec{B}}{|\vec{B}|}\times\left[\vec{J}\times\frac{\vec{B}}{|\vec{B}|}\right]\right|^2, 
        \label{eqn:eq1-e}\\
         p &=\rho \mathcal{R} \frac{T}{\tilde{\mu}},
        \label{eqn:eq1-f}
    \end{align}\label{eqn:sys_mhd_eq}
\end{subequations}
where $\frac{D}{Dt}=\frac{\partial}{\partial t}+\vec{v}\cdot\nabla$ is the total derivative. Equations in System~(\ref{eqn:sys_mhd_eq}) are the continuity equation, momentum equation, induction equation,  energy equation, and ideal gas equation, respectively.  In these expressions, $\vec{v}$ is the velocity, $\vec{B}$ is the magnetic field, $\rho$ is the mass density and $p$ is the thermal pressure, $\vec{J} = \left(\nabla \times  \vec{B} \right)/\mu_0$ is the current density; with $\mu_0$ the magnetic permeability, $\eta_A$ is the coefficient of ambipolar diffusion, $\gamma$ is the adiabatic index, $T$ is the temperature, $\mathcal{R}$ is the ideal gas constant, and $\tilde{\mu}$ is the mean atomic weight. As our focus shall be put on the effects of ambipolar diffusion, other non-ideal mechanisms, such as viscosity, radiative losses, thermal conduction, or the Hall effect are neglected. The ambipolar diffusion  is in many cases the most important non-ideal effect in partially ionized solar plasmas, although the Hall effect can be of importance in certain processes, such as reconnection or magneto-convection \citep[see, e.g.,][]{pandey2008,Threlfall2012,gonzalez2020}. In the middle and high chromosphere, as well as in solar prominences, the ambipolar diffusion is expected to be the dominant effect \citep[see details in, e.g.,][]{Khomenko2012,2018BallesterPartiallyIonizedPlasma}. Gravity is also dropped. 

We use the small perturbation theory to obtain a set of equations that describe the interplay between the MHD waves and the background shear flow. It is assumed that all the physical variables can be written as 
\begin{equation}
    f(\vec{r},t) = f_0(\vec{r}) + f_1(\vec{r}, t), 
\end{equation}
where $f_0$ is the equilibrium value and $f_1$ is a perturbation, which contains  the temporal dependence and is assumed to be much smaller than the equilibrium value, i.e.,  $|f_0| \gg |f_1|$. Hence, we linearize System~(\ref{eqn:sys_mhd_eq}), and the resulting linearized equations are
\begin{subequations}
    \begin{align}
        \left(\partial_t+Ax\partial_z\right)d&\:=-\left(\partial_xv_x+\partial_yv_y+\partial_zv_z\right),
        \label{eqn:cont_eq_lin_def}\\
        \left(\partial_t+Ax\partial_z\right)v_x&=-V_s^2\partial_xd+V_A^2(\partial_zb_x-\partial_xb_z),
        \label{eqn:momx_eq_lin_def}\\
        \left(\partial_t+Ax\partial_z\right)v_y&=-V_s^2\partial_yd+V_A^2(\partial_zb_y-\partial_yb_z),
        \label{eqn:momy_eq_lin_def}\\
        \left(\partial_t+Ax\partial_z\right)v_z&=-V_s^2\partial_zd-Av_x,
        \label{eqn:momz_eq_lin_def}\\
        \left(\partial_t+Ax\partial_z\right)b_x&=\partial_zv_x+\eta_A(\partial_z^2b_x-\partial_{zx}^2b_z),
        \label{eqn:indx_eq_lin_def}\\
        \left(\partial_t+Ax\partial_z\right)b_y&=\partial_zv_y+\eta_A(\partial_z^2b_y-\partial_{zy}^2b_z),
            \label{eqn:indy_eq_lin_def}
    \end{align}
    \label{eqn:sys_mhd_lin}
\end{subequations}
where we introduced the dimensionless perturbations of density, $d=\rho_1/\rho_0$, and magnetic field, $b_i=B_{1i}/B_0$, with $i=x,y,z$. To simplify the notation we dropped the subscript 1 from the components of the velocity perturbation, so that we simply write $v_i$ with $i=x,y,z$.   We note that the $z$-component of the linearized induction equation is not used and is conveniently replaced by the solenoidal constraint, namely
\begin{equation}
\partial_xb_x+\partial_yb_y+\partial_zb_z=0.\label{eqn:sol_eq_lin_def}
\end{equation}
Additionally, $V_A$ and $V_s$ represent the equilibrium Alfv\'en and sound speeds, respectively, and are defined by
\begin{align}
    V_A=\sqrt{\frac{B_0^2}{\mu\rho_0}}, \qquad V_s=\sqrt{\frac{\gamma p_0}{\rho_0}}.
    \label{eqn:sys_params_speeds}
\end{align}

 It should be noted that the source term due to ambipolar heating in Equation~(\ref{eqn:eq1-e}) drops as a result of the linearization, since it is a second-order term in the perturbations. Consequently, the internal energy equation becomes adiabatic in the linear approximation.

\subsection{Nonmodal analysis}

The usual asymptotic normal mode analysis is unsuited to generally study the interaction between MHD waves and shear flows, since the associated eigenvalue problem becomes non-self-adjoint due to the background flow  inhomogeneity \citep{1993TrefethenNonmodalOperators, 1993ChagelishviliETransMechanism}. This has led to the use of so-called nonmodal
approach \citep{1999MahajanNonmodalAnalysis, 2001BodoSpatialWaveTransformation,2012CamporealeNonmodalLinearTheory}, in which one considers the temporal evolution of the spatial Fourier harmonics of the perturbations, but without any spectral expansion in time.

Following the standard method of the nonmodal approach, we introduce the change of variables,
\begin{align}
    x' = x, \qquad y' = y, \qquad z' = z-Axt, \qquad t' = t, 
\end{align}
in order to transform the spatial inhomogeneity into a temporal one. In the following expressions, the prime symbol $'$ on the new variables is dropped for simplicity. We continue by taking the spatial Fourier transform of the perturbations. Hence, any perturbation, generally represented  as $F$, is expressed as,
\begin{equation}
    F(x, y, z, t) = \int\tilde{F}(k_x, k_y, k_z, t) e^{i(k_xx+k_yy+k_zz)}dk_xdk_ydk_z, 
\end{equation}
where $\tilde{F}$ is the Fourier transform of $F$, with $k_x$, $k_y$, and $k_z$ the wavenumbers  in the $x$, $y$, and $z$ directions, respectively. This allows us to rewrite the set of equations in System~(\ref{eqn:sys_mhd_lin}) in terms of the Fourier transforms of the perturbations, where all spatial derivatives are replaced by products with the respective wavenumbers. Introducing dimensionless variables, the new set of equations in Fourier space is,
\begin{subequations}
    \begin{align}
        D^{(1)} &= K_xu_x + K_yu_y + u_z,
        \label{eqn:cont_eq_F_trans_adim}\\
        u_x^{(1)}&= -\varepsilon^2K_xD + \left(1+K_x^2\right)b_x + K_xK_yb_y, 
        \label{eqn:momx_eq_F_trans_adim}\\
        u_y^{(1)} &= -\varepsilon^2K_yD + K_xK_yb_x + \left(1+K_y^2\right)b_y,
        \label{eqn:momy_eq_F_trans_adim}\\
        u_z^{(1)} &= -\varepsilon^2D -Ru_x,
        \label{eqn:momz_eq_F_trans_adim}\\
        b_x^{(1)} &= -u_x - \tilde{\eta}_A\left(1+K_x^2\right)b_x - \tilde{\eta}_A K_xK_yb_y,
        \label{eqn:indx_eq_F_trans_adim}\\
        b_y^{(1)} &= -u_y - \tilde{\eta}_A K_xK_yb_x - \tilde{\eta}_A\left(1+K_y^2\right) b_y,
        \label{eqn:indy_eq_F_trans_adim}
    \end{align}
    \label{eqn:sys_mhd_dimless}
\end{subequations}
where $f^{(n)}$ denotes the $n$th-order temporal derivative of $f$ and we used Equation~(\ref{eqn:sol_eq_lin_def}) to eliminate $b_z$. The dimensionless variables and coefficients are defined as,
\begin{subequations}
    \begin{align}
        D&=i\tilde{d}, \qquad b_x = i \tilde{b}_x, \qquad   b_y= i \tilde{b}_y,\\
        u_x &= \frac{\tilde{v}_x}{V_A}, \qquad  u_y= \frac{\tilde{v}_y}{V_A},  \qquad u_z= \frac{\tilde{v}_z}{V_A},\\
        K_x &= \frac{k_x}{k_z}-R\tau=K_{x0}-R\tau, \label{eq:kx} \qquad  K_y=\frac{k_y}{k_z},\\      R&=\frac{A}{V_A k_z}, \qquad         \varepsilon=\frac{V_s}{V_A}, \qquad       \tilde{\eta}_A = \frac{k_z\eta_A}{V_A}.
        \end{align}
    \label{eqn:sys_dimensionless_system_paramenters}
\end{subequations}
We note that the velocities are all normalized to the Alfv\'en speed. The new temporal variable, $\tau$, is normalized to the Alfv\'en time, so that $\tau = V_Ak_zt$. In addition, $R$ measures the normalized strength of the velocity shear and $\varepsilon$ represents, essentially, the plasma $\beta$, i.e., the importance of the thermal pressure with respect to the magnetic pressure. Importantly, we note that in this formalism  $K_x$ is a function of time. If $K_{x0} > 0$ and for a sufficiently long time, $K_x$ will change sign. This can be understood as a  reflection of the wave caused by the shear flow.

The set of Equations~(\ref{eqn:sys_mhd_dimless}) may further be reduced into a system of three intercoupled second order ordinary differential equations (ODEs) by introducing the new variable $\psi = D + K_xb_x + K_yb_y$ and expressing $u_x$, $u_y$ and $u_z$ and their temporal derivatives in terms of the magnetic field perturbations, namely
\begin{subequations}
    \begin{align}
        \begin{split} \label{eqn:derivació_eq_11a_def}
               \psi^{(2)} + \omega_1^2\psi &= (C_{12}+ \gamma_1)b_x + (C_{13}+ \gamma_2)b_y +\\
             & + J_1b_x^{(1)} + J_2b_y^{(1)},
        \end{split}\\
        \begin{split} \label{eqn:derivació_eq_11b_def}
             b_x^{(2)} + \left(\omega_2^2 + \gamma_3\right)b_x &= C_{12}\psi + (C_{23}+ \gamma_4)b_y +\\
             & + J_3b_x^{(1)} + J_4b_y^{(1)},
        \end{split}\\
        \begin{split} \label{eqn:derivació_eq_11c_def}
            b_y^{(2)} + \omega_3^2b_y &= C_{13}\psi + (C_{23} + \gamma_4)b_x + \\
             & + J_4b_x^{(1)} + J_5b_y^{(1)} ,
        \end{split} 
    \end{align}   
    \label{eqn:sys_eqs_11_diffusion}
\end{subequations}        
where the following auxiliary notation has been introduced: the uncoupled oscillation eigenfrequencies, $\omega_1$, $\omega_2$, and $\omega_3$, namely
\begin{subequations}
    \begin{align}
    \omega_1 &= \varepsilon, \\
        \omega_2 &= \sqrt{1+\left(1+\varepsilon^2\right)K_x^2}, \\
        \omega_3 &= \sqrt{1+\left(1+\varepsilon^2\right)K_y^2},
    \end{align}
    \label{eqn:sys_aux_coefs_freqs}
\end{subequations}
the ideal coupling coefficients, $C_{ij}$, namely
\begin{align}
        C_{12} = \varepsilon^2K_x, \qquad
        C_{13} = \varepsilon^2K_y, \qquad
        C_{23} = -\left(1+\varepsilon^2\right)K_xK_y,
        \label{eqn:sys_aux_coefs_coupling_coefs}
\end{align}
and the non-ideal coefficients, $\gamma_i$ and $J_i$,  associated with coupling and dissipation owing to ambipolar diffusion, namely
\begin{subequations}
    \begin{align}
        \gamma_1&= \tilde{\eta}_A R\left(2+4K_x^2+K_y^2\right), \qquad
        &\gamma_2&= 3\tilde{\eta}_A RK_xK_y, \\
        \gamma_3&= -2\tilde{\eta}_A RK_x, \qquad
        &\gamma_4&= \tilde{\eta}_A R K_y, \\
        J_1 &= -K_x\tilde{\eta}_A\left(1+K_x^2+K_y^2\right), \qquad
        &J_4 &= -\tilde{\eta}_A K_xK_y,\\
        J_2 &= -K_y\tilde{\eta}_A\left(1+K_x^2+K_y^2\right), \qquad
        &J_5 &= -\tilde{\eta}_A\left(1+K_y^2\right),\\
        J_3& = -\tilde{\eta}_A\left(1+K_x^2\right).
    \end{align}
    \label{eqn:sys_aux_coefs_diff_coefs_2} 
\end{subequations}
The process to arrive at the set of Equations~(\ref{eqn:sys_eqs_11_diffusion}) is summarized as follows.
To obtain Equations~\ref{eqn:derivació_eq_11b_def} and \ref{eqn:derivació_eq_11c_def}, we begin by differentiating Equations~(\ref{eqn:indx_eq_F_trans_adim}) and (\ref{eqn:indy_eq_F_trans_adim}) with respect of time and then substituting the expressions for $u_x^{(1)}$ and $u_y^{(1)}$ into Equations~(\ref{eqn:momx_eq_F_trans_adim}) and (\ref{eqn:momy_eq_F_trans_adim}), respectively. Subsequently, we replace $D$ by the new variable $\psi$. 
To arrive at Equation~(\ref{eqn:derivació_eq_11a_def}), we start by substituting $u_x$ and $u_y$ from Equations~(\ref{eqn:indx_eq_F_trans_adim}) and (\ref{eqn:indy_eq_F_trans_adim}) into Equation~(\ref{eqn:cont_eq_F_trans_adim}). The resulting expression is then differentiated with respect of time and $u_z^{(1)}$ is eliminated using Equation~(\ref{eqn:momz_eq_F_trans_adim}). Later, $u_x$ is again substituted using Equation~(\ref{eqn:indx_eq_F_trans_adim}). Finally,  we again replace $D$ by the new variable $\psi$.

System (\ref{eqn:sys_eqs_11_diffusion}) describes coupled linear oscillations with three degrees of freedom in a plasma with ambipolar diffusion and a shear flow. Owing to presence of shear in the flow, i.e., when $R\neq0$, some of the coefficients have time-dependent behavior.  This set of equations is quite general and describes all the linear MHD modes, their mutual transformations and interaction with the shear flow, and the damping owing to ambipolar diffusion. Generally, the linear MHD modes are the Alfv\'en wave, AW, the fast magnetosonic wave, FMW, and the slow magnetosonic wave, SMW.

To compare with the previous results published in \citet{1998PoedtsSFWaveCoupling} corresponding to the case without diffusion, we set $\tilde{\eta}_A = 0$ and System (\ref{eqn:sys_eqs_11_diffusion}) simplifies to,
\begin{subequations}
    \begin{align}
       \psi^{(2)} + \omega_1^2\psi =& C_{12}b_x + C_{13}b_y, \label{eqn:derivació_eq_11a_ideal} \\
        b_x^{(2)} + \omega_2^2b_x =& C_{12}\psi + C_{23}b_y,
        \label{eqn:derivació_eq_11b_ideal} \\
        b_y^{(2)} + \omega_3^2b_y =& C_{13}\psi + C_{23}b_x.
        \label{eqn:derivació_eq_11c_ideal} 
    \end{align}
    \label{eqn:sys_eqs_11_ideal}
\end{subequations}  
This set of three equations consistently reverts to that given in \citet{1998PoedtsSFWaveCoupling}.
 
The Wolfram Mathematica environment has been used to obtain numerical solutions of System (\ref{eqn:sys_eqs_11_diffusion}) for prescribed initial conditions and a given set of model parameters.  We used the module \verb|NDSolve| to perform the integration and chose a small enough time step to avoid introducing relevant  errors. This has been determined by convergence tests. The numerical solution  is stored as an \verb|InterpolatingFunction| for each perturbation, which allows their manipulation and representation as if they were analytical functions.

\subsection{Considerations about wave coupling} \label{chap:freqs-E}

Before exploring the numerical solutions of the system of Equations~(\ref{eqn:sys_eqs_11_diffusion}), here we enumerate some considerations that can be useful to understand the results. These are based on previous findings discussed in, e.g, \citet{1996ChagelishviliCoupling&LinearTransformation,1997ChagelishviliMHDWaves,1998PoedtsSFWaveCoupling,2004GogoberidzeLinearCoupling}, among others.

We begin by studying how the coupling of the  MHD modes can occur in the  case where only two modes are considered  without dissipation \citep{1981Morse}. In this situation, we can describe the system mathematically as a pair of coupled linear oscillators, which leads to the following set of second-order ODEs for the variables $F_1$ and $F_2$,
\begin{equation}
    F_1^{(2)} + \omega_1^2F_1 + CF_2 = 0, \qquad 
    F_2^{(2)} + \omega_2^2F_2 + CF_1 = 0,
    \label{eqn:sys_canonical_coupled_os3cs}
\end{equation}
\noindent where $\omega_1$ and $\omega_2$ are the oscillator eigenfrequencies, and $C$ is the coupling coefficient. If one allows both oscillators to move simultaneously, that will usually result in a non-periodic motion \citep{1981Morse}. However, if the eigenfrequencies and coupling coefficient are constant, the general solution of the system can always be described as a combination of the normal modes. The normal frequencies of the coupled oscillations, $\Omega_{\pm}$, can be determined with
\begin{equation}
    \Omega_{\pm}^2=\frac{1}{2} \left[\left(\omega_1^2+\omega_2^2\right)\pm \sqrt{\left(\omega_1^2-\omega_2^2\right)+4C^2}\right].
    \label{eqn:freqs_fonamentals}
\end{equation}
Note that the value of $\Omega_-$ is smaller than either $\omega_1$ or $\omega_2$, while $\Omega_+$ is larger than both eigenfrequencies, meaning that the coupling will always spread apart the natural frequencies.

As one can see in the eigenfrequencies and coupling coefficients of our system, in Equations (\ref{eqn:sys_aux_coefs_freqs}) and (\ref{eqn:sys_aux_coefs_coupling_coefs}), they depend on time through  $K_x$, thus making the canonical theory of coupled oscillation invalid. However, when $R\ll1$ the system evolves slowly enough and the standard theory of coupled oscillations may be a useful reference in the  interpretation of the actual coupling. In the case where the three MHD waves are considered, we can use the eigenfrequencies of the ideal waves as a reference. In our normalized notation these frequencies are,
\begin{subequations}
    \begin{align}
        \Omega_{F,S}^2 &= \frac{1}{2}K^2\left(1+\varepsilon^2\right)\left(1\pm\sqrt{1-\frac{4\varepsilon^2}{K^2\left(1+\varepsilon^2\right)^2}}\right),
        \label{eqn:freq_FMW_SMW}\\
        \Omega_A&=1,
        \label{eqn:freq_AW}
    \end{align}
    \label{eqn:sys_eigenfreqs_MHD_waves}
\end{subequations}
where $K^2 = 1 + K_x^2 + K_y^2$ represents the modulus of the wave vector squared and the subscripts $F$, $S$, and $A$ correspond to the FMW, the SMW, and the AW, respectively. Strictly speaking, these frequencies are not valid in our setup due to  the temporal variation of some parameters and the ambipolar diffusion. The difference between coupled and uncoupled frequencies is presumably  small. This happens when the temporal variation of the wave number $K_x(\tau)$ is slow enough, which is realized when $\partial_\tau\Omega_i\ll\Omega_i^2$, with $i=F \text{, } S\text{, } A$. 

When the eigenfrequencies and/or coupling coefficient vary slowly in time, the system displays mutual transformation with the corresponding energy transfer between the three modes \citep{1996ChagelishviliCoupling&LinearTransformation,1997ChagelishviliMHDWaves,2013KotkinCollectionProblems}. Similar problems have been studied in the past and two necessary conditions were found for the effectiveness of the energy exchange between  weakly coupled oscillators \citep[see][]{2013KotkinCollectionProblems}:

\begin{enumerate}

    \item There should exist a so called degeneration region (DR) where $|\Omega_i^2-\Omega_j^2|\leq|C_{ij}(\tau)|$. Here $\Omega_{i,j}$ represents the frequencies of two oscillation modes that couple and $C_{ij}$ the corresponding coupling coefficient. For the weak coupling case, this implies that $\Omega_i\simeq \Omega_j$, thus the maximum energy exchange between modes happens when their frequencies are approximately the same.

    \item The DR has to be crossed slowly. In other words, the shear-modified frequencies have to vary slowly enough so that $|\Omega_{i,j}^{(1)}(\tau)|\ll|C_{ij}(\tau)|$. This condition readily holds in the DR when we consider a low shear flow with $R\ll1$.
    
\end{enumerate}

Hence, from the first condition we learn that the DR has to be found in the neighborhood of the time that minimizes the difference between frequencies, so that $|K_x(\tau)|\leq 1$ \citep{1998PoedtsSFWaveCoupling}. This characteristic time where $K_x(\tau)=0$ is denoted as $\tau_*$ and can be calculated as
\begin{align}
    \tau_*=\frac{K_{x0}}{R}. \label{eq:taustar}
\end{align}
From this expression we can see that the temporal scale of energy exchange between background flow and wave modes is of the order $\sim R^{-1}$. Furthermore, $\tau_*$ can also be interpreted as the time at which the effective reflection of the wave propagation in the $x$-direction occurs. The sheared background flow attempts to change the direction of wave propagation across the flow, resulting in the temporal dependence of $K_x(\tau)$. When $\tau=\tau_*$ and for $K_{x0} > 0$ initially, the value of $K_x(\tau)$ reaches zero and begins to increase negatively, resulting in the effective reflection of the wave. When this phenomenon of effective wave reflection happens, the shear flow could be seen from an external observer as a driver of the waves.

On the other hand, from the second condition it can be shown that the temporal scale of resonant interaction, this is the time scale of energy exchange between wave modes, is of the order $\sim R^{-1+1/R}$ \citep[see details in][]{2004GogoberidzeLinearCoupling}. If we consider the low shear case ($R\ll1$), we reach the conclusion that the resonant interaction of waves is a faster process than the energy exchange between the shear flow and wave modes \citep{2004GogoberidzeLinearCoupling}.

Additionally, it is important to note that the wave transformation is subject to temporal restrictions but not to spatial ones. Thus, transformation of waves in shear flows may potentially happen throughout the medium, but only at the specific times when the DR is realized locally \citep{1997ChagelishviliMHDWaves}.

Let us clarify the necessary conditions on the system parameters to achieve the DR and, therefore, energy exchange between MHD waves and shear flow and mutual transformation between the MHD waves can happen:

\begin{itemize}
    
    \item Mutual transformation between FMW and AW, 
    \begin{displaymath}
        \Omega_F^2\simeq\Omega_A^2, \qquad \varepsilon<1, \qquad K_y\ll1.
    \end{displaymath}

    \item Mutual transformation between FMW and SMW:
\begin{displaymath}
        \Omega_F^2\simeq\Omega_S^2, \qquad \varepsilon \simeq 1.
    \end{displaymath}

    \item Mutual transformation between AW and SMW: 
    \begin{displaymath}
        \Omega_A^2\simeq\Omega_S^2, \qquad \varepsilon > 1.
    \end{displaymath}

    \item Mutual transformation between the three modes: 
        \begin{displaymath}
        \Omega_F^2\simeq\Omega_A^2\simeq\Omega_S^2, \qquad \varepsilon\simeq1, \qquad K_y\ll1.
\end{displaymath}

\end{itemize}

A particular case of the mutual transformation between the three modes is the so called beat regime, which is realized when $R\ll K_{x0}\ll 1$.   This  leads to an inherently efficient coupling \citep[see][]{1997RogavaCouplingHydrodinamicalWaves,1998PoedtsSFWaveCoupling}. 

The analysis of the energy associated to the  wave modes can be  useful to understand the coupling. Following the process described in \citet{walker2005}, we have derived a conservation equation for the energy of the linear perturbations, namely
\begin{equation}
    \label{eqn:Energy_conservation}
    \partial_t U + \vec{V}_0\cdot\nabla U + \nabla\cdot\vec{\Pi} = H_{\rm flow}+H_{\rm diff.},
\end{equation}
where
\begin{equation}
    \label{eqn:Energy_density}
    U = \frac{1}{2}\rho_0|\vec{v}_1|^2+\frac{1}{2\mu_0}|\vec{B}_1|^2+\frac{1}{2\gamma p_0}p_1^2,
\end{equation}
is the energy density, with the term $\vec{V}_0\cdot\nabla U$ accounting for the advection of the energy density by the mass flow,
\begin{equation}
    \label{eqn:Energy_flux}
    \vec{\Pi}= \left(p_1+\frac{\vec{B}_0\cdot \vec{B}_1}{\mu_0}\right)\vec{v}_1-\frac{\vec{v}_1\cdot \vec{B}_1}{\mu_0}\vec{B}_0,
\end{equation}
is the waves energy flux,
\begin{equation}
    \label{eqn:Energy_exchange_w_flow}
    H_{\rm flow}=\vec{B}_1\cdot\left[(\vec{B}_1\cdot\nabla)\vec{V}_0\right]-\vec{v}_1\cdot\left[(\vec{v}_1\cdot\nabla)\vec{V}_0\right],
\end{equation}
is a source/sink term due to the energy exchange between the waves and the shear flow, and
\begin{equation}
    \label{eqn:Energy_diffusion_term}
    H_{\rm diff.}=\frac{\eta_A}{|\vec{B}_0|^2}\vec{B}_1\cdot\left\{\left(\left(\nabla\times \vec{B}_1\right)\times \vec{B}_0\right)\times \vec{B}_0\right\},
\end{equation}
is the energy dissipation term due to ambipolar diffusion. The expressions in Equations (\ref{eqn:Energy_density})--(\ref{eqn:Energy_diffusion_term}) are written in a general form. It is  instructive to obtain the  expression of $H_{\rm flow}$ in the specific configuration of our background flow with  $\vec{V}_0=Ax\hat{z}$, namely
\begin{equation}
     H_{\rm flow} = A\left(B_{1x}B_{1z}-v_{x}v_{z}\right). \label{eq:hflow}
\end{equation}
First, it is worth noting that $H_{\rm flow}$ can be either positive or negative, depending on the phase relations between the oscillating perturbations, so that it can act as either an energy source term (energy goes from the flow to the waves) or an energy sink term (energy goes from the waves to the flow).  Equation~(\ref{eq:hflow}) also shows that the energy exchange between waves and flow is proportional to the shear strength, $A$. This means that the larger the shear, the more important this energy exchange is. In addition, $H_{\rm flow}$ involves the $x$- and $z$-components of the velocity and magnetic field perturbations, but not the $y$-components. As a consequence of this, pure Alfv\'en waves, which are polarized in the $y$-direction, are unable the exchange energy with the flow. In relation to this comment, \citet{1996ChagelishvilAmpVortexDistSF} showed that perturbations in thermal or magnetic pressure are required for the energy exchange with the flow to happen. Hence, only the waves capable of generating such perturbations can exchange energy with the flow. Consequently, as linear Alfv\'en waves are incompressible and do not perturb the pressure field, they cannot exchange energy with the background flow by themselves via shear-induced processes  \citep{1997ChagelishviliMHDWaves}.

Ambipolar diffusion appears in the wave energy equation as a sink term (Equation~\ref{eqn:Energy_diffusion_term}). Since energy is a conserved quantity, the energy lost by the wave due to ambipolar diffusion must be deposited into the plasma and converted into internal energy. However, as previously discussed, the internal energy equation becomes adiabatic under linearization, because the ambipolar heating term is quadratic in the perturbations and therefore vanishes in the linear regime. This limitation is common to all studies addressing the dissipation of linear waves. Although the present linear analysis cannot capture the actual heating of the background plasma, the rate at which wave energy is dissipated must, by conservation of energy, match the rate of internal energy increase in the full nonlinear system. This point has been discussed in, e.g., \citet{soler2016,soler2017,cally2023}.

Now, we denote by $E$ the energy density in dimensionless units. It can be decomposed as:
\begin{equation}
    E=E_k+E_m+E_c,  \label{eqn:Total_Energy_density}
\end{equation}
where $E_k$, $E_m$, and $E_c$ are the dimensionless kinetic, magnetic and compression energy densities, respectively, given by
\begin{subequations}
    \begin{align}
        E_k&=\frac{1}{2}\left(|u_x|^2+|u_y|^2+|u_z|^2\right), 
        \label{eqn:kinetic_Energy_density}\\
        E_m&=\frac{1}{2}\left(|b_x|^2+|b_y|^2+|b_z|^2\right),  
        \label{eqn:magnetic_Energy_density}\\
        E_c&=\frac{\varepsilon^2 }{2}|D|^2.
        \label{eqn:compression_Energy_density}
    \end{align}
    \label{eqn:sys_Energy_density}
\end{subequations}
In turn, the dimensionless expression of the energy exchange term with the shear flow  is
\begin{equation}
     \bar{H}_{\rm flow} = -R\left(b_{x}b_{z}+u_{x}u_{z}\right), 
\end{equation}
which clearly shows that this term is proportional to the shear parameter, $R$, so that the previous statement that the temporal scale of energy exchange between background flow and wave modes is of the order $\sim R^{-1}$ is confirmed. In a ideal, low-shear scenario, the temporal evolution of the energy of the modes because of the interaction with the flow proceeds slowly. This behavior implies that the energy density of each oscillation mode should normally evolve proportionally to its own dispersion curve ($E_i\propto\Omega_i$) due to the exchange of energy between the mode and the flow. As already discussed, this exchange can work in both directions, meaning that the mode can either extract energy from the flow, and thus the mode energy increases, or give energy to the flow, and thus the mode energy decreases. This holds true at any time, except when the system is in the DR, where effective transformations between modes happens. The mode-mode transformations within the DR are only partial and part of the energy is still exchanged with the background flow, which results in the energy evolution ceasing to be proportional to the frequencies. 

In summary, the energy evolution of each mode is proportional to its frequency until it reaches the DR, where effective energy exchange between mode takes place and the proportionality is broken. Afterwards, when the system exits the DR, the proportionality between the evolution of the energies and the oscillation frequencies is restored. In the presence of ambipolar diffusion, however, we expect the proportionally between wave energy and wave frequency to hold no longer, due to the energy dissipation.

\section{Ideal FMW-AW transformation} \label{ND F-ATr}

Although we have presented the governing Equations~(\ref{eqn:sys_eqs_11_diffusion}) in their most general form, which describes the transformations between the three types of MHD waves, in this paper we restrict ourselves to the FMW-AW transformation by setting $\varepsilon=0$. This removes the SMW from the scene, suppressing all the $\psi$ terms from Equations~(\ref{eqn:sys_eqs_11_diffusion}). The coupling with the SMW will be investigated in the forthcoming second part of this work.

As an initial case we consider an ideal system where the effect of ambipolar diffusion is neglected, so that $\tilde{\eta}_A = 0$. Thus, further reducing Equations~(\ref{eqn:sys_eqs_11_diffusion}) to the following pair of coupled second-order ODEs\footnote{We have to note the presence of some typographical errors in the equivalent equations given in \citet{1998PoedtsSFWaveCoupling}, see their set of Equations~(14). We provide the correct expressions.}:
\begin{subequations}
    \begin{align}
        b_x^{(2)} + \left(1+K_x^2\right)b_x + K_xK_yb_y = 0,
        \label{eqn:derivació_eq_14a_ideal}\\
        b_y^{(2)} + \left(1+K_y^2\right)b_y + K_xK_yb_x = 0,
        \label{eqn:derivació_eq_14b_ideal}
    \end{align}
    \label{eqn:sys_Trs_FMW_AW_ideal}
\end{subequations}
which describe the interaction between compressional pressureless FMW and shear incompressible AW \citep[see][]{1996RogavaAlfvenWaves,1998PoedtsSFWaveCoupling}. This system is mathematically equivalent to a pair of pendulums connected by a spring with a varying stiffness coefficient, and the length of one of these pendulums also varies in time, since $K_x$ is a function of time.
The dimensionless eigenfrequencies of the oscillators are $\omega_F^2=1+K_x^2$ and $\omega_A^2=1+K_y^2$, and the coupling coefficient is $C_{FA} = K_xK_y$. The fundamental frequencies of these oscillations may be calculating using Equations (\ref{eqn:freqs_fonamentals})
\begin{equation}
    \Omega_F^2\equiv\Omega_+^2=1+K_x^2+K_y^2,\qquad
    \Omega_A^2\equiv\Omega_-^2=1,
    \label{eqn:sys_freqs_TRS_FMW_AW_ideal}
\end{equation}
and correspond, respectively, to the FMW and AW frequencies.  It is easy to identify that $\Omega_F$ depends on $\tau$ through $K_x$. An interesting feature of the shear-induced wave coupling in this scenario is that its efficiency depends mainly on the values of $K_{x0}$ and $K_y$, which determine the initial direction of the wavevector, and not explicitly on the inherent physical characteristics of the background flow. The energy of the system is given by the following relation,
\begin{equation}
    E =\frac{1}{2}\left(|u_x|^2+|u_y|^2+|u_z|^2+|b_x|^2+|b_y|^2+|b_z|^2\right).
    \label{eqn:Energy_F_A}
\end{equation}
Furthermore, in this case where only the FMW and the AW are present, and  the magnetic field and the background flow are both directed along the $z$-axis, the two waves are polarized perpendicularly to each other and we can estimate their respective energy contributions, namely
\begin{subequations}
\begin{align}
    E_{F}&=\frac{1}{2}\left(|u_x|^2+|u_z|^2+|b_x|^2+|b_z|^2\right),\\
    E_{A}&=\frac{1}{2}\left(|u_y|^2+|b_y|^2\right).
\end{align}
    \label{eqn:Energy_F_A_contributions}
\end{subequations}

\begin{figure}
    \includegraphics[width=
    \hsize]{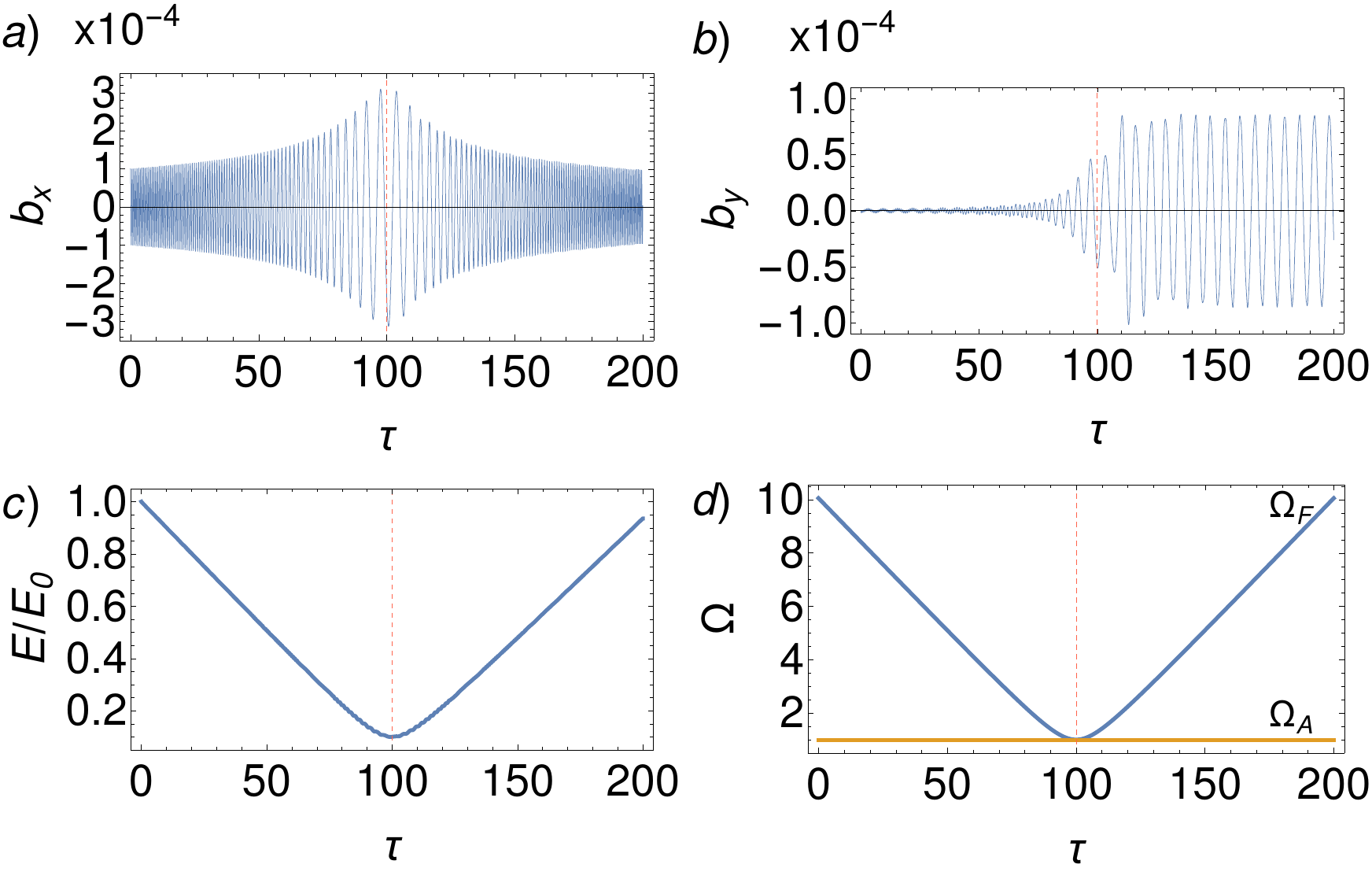}
    \caption{Ideal FMW-AW transformation with parameters $K_{x0}=10$, $R=0.1$, and $K_y=0.1$. Temporal evolution of a) $b_x(\tau)$,  b) $b_y(\tau)$, c) the normalized total energy, $E(\tau)/E(0)$, and d) wave frequencies (dispersion curves) of the FMW and the AW. In panel d) the blue and orange lines denote the fast frequency and the Alfv\'en frequency, respectively. In all panels, the red dashed line marks $\tau_*$. }
    \label{fig:fmw_aw_Poedts_id_by_E_freq}
\end{figure}

As an initial check, we show in Figure \ref{fig:fmw_aw_Poedts_id_by_E_freq} the results of the numerical solution of  Equations (\ref{eqn:sys_Trs_FMW_AW_ideal}) corresponding to one of the cases studied in \citet{1998PoedtsSFWaveCoupling} with parameters $K_{x0}=10$, $K_y=0.1$, and $R=0.1$. We show the representations for $b_x(\tau)$, $b_y(\tau)$, the normalized $E(\tau)$, and frequencies in Figure~\ref{fig:fmw_aw_Poedts_id_by_E_freq}, although the plots of $b_y(\tau)$ and the normalized $E(\tau)$ are the only ones shown in \citet{1998PoedtsSFWaveCoupling}. An initially pure FMW, excited using the  initial conditions $b_x(0)=10^{-4}$ and $b_y(0)=b_x'(0)=b_y'(0)=0$, transforms partially into an AW around the time $\tau_*=100$. That can be seen clearly in Figure \ref{fig:fmw_aw_Poedts_id_by_E_freq}(d), where the  frequency $\Omega_F$ shifts from an initially high value to a lower one corresponding to $\Omega_A$. 

In addition to the FMW transferring part of its energy to the AW, the evolution of the wave energy in Figure~\ref{fig:fmw_aw_Poedts_id_by_E_freq}c) evidences how the FMW exchanges energy with the flow. First, the wave energy decreases for $\tau < \tau_*$ owing to a net flux of energy from the FMW to the background flow. Physically, this is caused by the shear flow attempting to reverse the direction of the FMW propagation. Then, $K_x$ changes sign when $\tau=\tau_*$, producing the effective reflection of the FMW. For $\tau>\tau_*$ the wave energy increases again, since now it is the flow that provides a net energy input to the FMW. In other words: the flow acts as a driver for the FMW. In the present linear model, the FMW energy increase happens indefinitely in time, because the stationary background flow is an infinite reservoir of energy. The time interval considered in Figure~\ref{fig:fmw_aw_Poedts_id_by_E_freq} is twice the value of $\tau_*$, which makes the wave energy evolution to appear roughly symmetric with respect to $\tau=\tau_*$. However, the wave energy would continue to increase for later times. 

Comparing our results to the ones presented in Figure~2 from \citet{1998PoedtsSFWaveCoupling}, we can see that they are very similar to each other. It is easy to identify that the evolution of $b_y(\tau)$ from a FMW to an AW happens at the same $\tau$ and the final amplitude of the wave is the same. On the other hand it seems that our system ends up with slightly higher energy. Why this happens is unclear to us. It may be due to slight differences in the definition of the problem or some numerical inaccuracy in the solution of \citet{1998PoedtsSFWaveCoupling}. We have been concerned to avoid numerical errors in the solutions. Nevertheless, comparing Figures \ref{fig:fmw_aw_Poedts_id_by_E_freq}(b) and \ref{fig:fmw_aw_Poedts_id_by_E_freq}(c) is obvious that outside of the DR, the energy is proportional to the fast frequency, $\Omega_F$, as it is expected in the low shear case, assuring that our numerical method solves the problem correctly. 

A peculiarity of such a system where only the FMW is initially generated and the evolution proceeds slowly, is that the energy transferred to the AW can be estimated from the change in the slope of the energy evolution (see Figure \ref{fig:fmw_aw_Poedts_id_by_E_freq}(d)). In this situation, as it has been commented earlier, the total energy of the system is proportional to the frequency of the FMW, $E=\lambda\Omega_{F}$, where $\lambda$ determines the slope of the energy. As the incompressible AW cannot directly interact with the flow, the slope of energy evolution after $\tau = \tau_*$ changes proportionally to the energy transmitted to the AW. The ratio of the slopes absolute value after and before $\tau=\tau_*$, computed as $\lambda_a/\lambda_b$, represents the fraction of initial energy that the FMW has preserved after the interaction with the AW, allowing us to approximately calculate the energy transferred to the AW as $E_{A}\approx E(\tau_*)\left(1-\frac{\lambda_a}{\lambda_b}\right)$. In the case presented in Figure~\ref{fig:fmw_aw_Poedts_id_by_E_freq}, the energy transmitted to the AW is around a $7.2\%$ the total wave energy in the system at $\tau=\tau_*$, which corresponds to about $0.7\%$ of the  initial wave energy. Nevertheless, it is important to note that  this provides a rough approximation of the  transmitted energy and is only valid under slow evolution.

We will use this example case as the reference for the rest of the studies performed in this work, namely a system with an initially pure FMW that partially transform into an AW nearby the time $\tau_*$, obtained using the initial conditions $b_x(0)\neq 0$ and $b_y(0)=b_x'(0)=b_y'(0)=0$.

\subsection{Dependence on $K_{x0}$} \label{Vars Kx0}

We  proceed to study how the various  parameters affect the solution, beginning with $K_{x0}$. We consider a system with $K_y=0.1$ and $R=0.1$, while we vary $K_{x0}$. All the considered cases meet the conditions presented in Section \ref{chap:freqs-E} for an effective coupling between the FMW and AW.

\begin{figure*}[htpb]
    \centering
    \includegraphics[width=0.9\hsize]{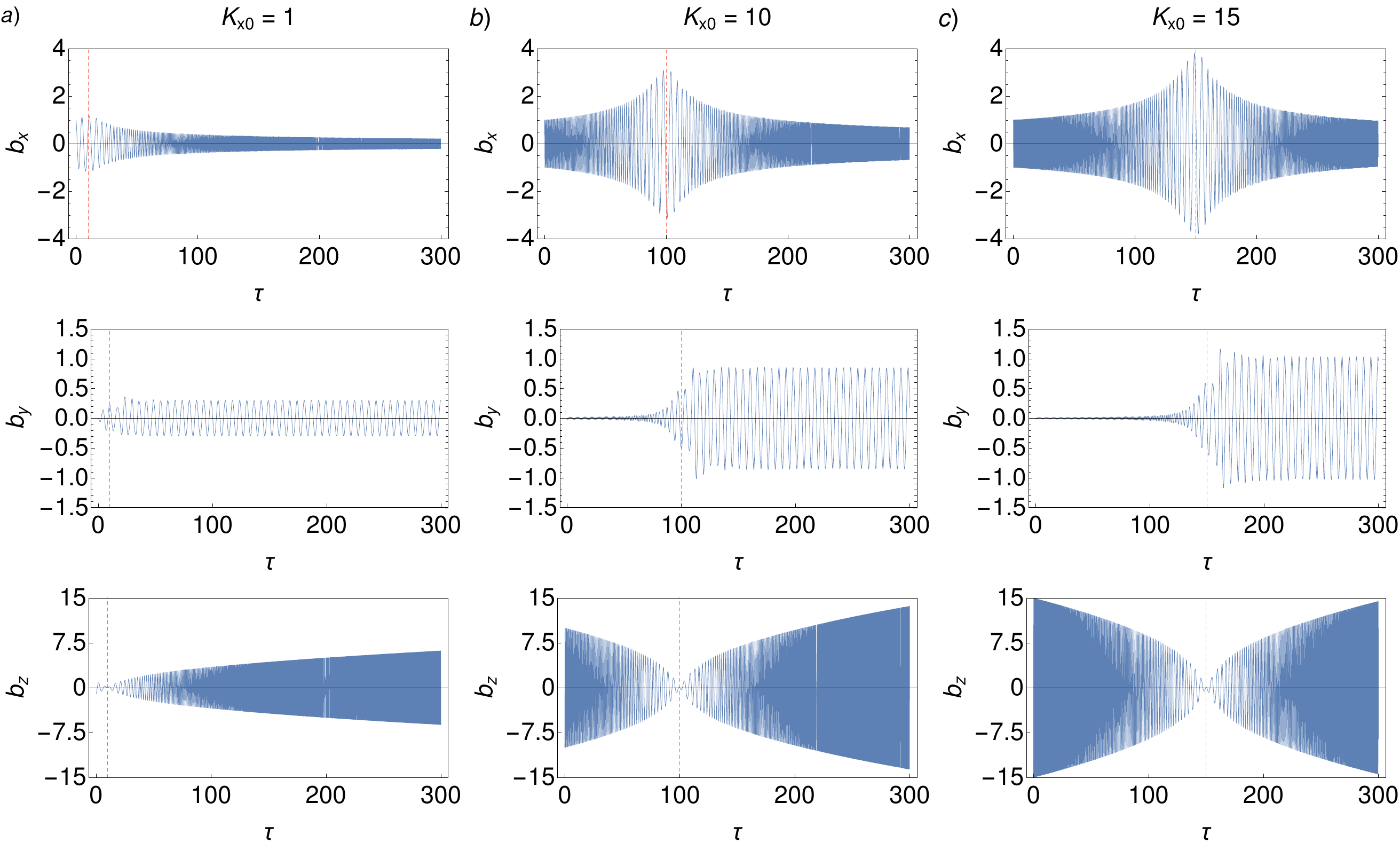}
    \caption{Ideal FMW-AW transformation with $R=0.1$ and $K_y=0.1$ for a) $K_{x0}=1$, b) $K_{x0}=10$, and c) $K_{x0}=15$. From top to bottom it is displayed the temporal evolutions of $b_x(\tau)$,  associated with the FMW, $b_y(\tau)$, associated with the AW, and $b_z(\tau)$,  related to the total pressure perturbation, respectively. The red dashed line marks $\tau_*$ in all plots.}
    \label{fig:Ideal_case_bx_by_Kx0_variation}
\end{figure*}

Figure~\ref{fig:Ideal_case_bx_by_Kx0_variation} shows the results of the temporal evolution of the magnetic field perturbations  for $K_{x0} =$~1, 10 and 15. In all cases, there is a partial transformation of the initially pure FMW into an AW around the time $\tau=\tau_*$, which depends on $K_{x0}$ (Equation~(\ref{eq:taustar})). For $\tau < \tau_*$, the $b_x$ and $b_y$ components of the magnetic field experience an amplification, while the amplitude of  $b_z$  decreases and becomes zero at $\tau=\tau_*$. As the total pressure perturbation  is proportional to $b_z$, this indicates that the total pressure perturbation becomes zero just at $\tau=\tau_*$, when the mode conversion is most efficient. This can be understood by the fact that the AW generates no pressure perturbations. Conversely, for $\tau>\tau_*$ the amplitudes of $b_x$ and $b_z$ progressively decrease and increase, respectively, while that of $b_y$, associated to the AW, remains constant since there is no further energy input to the AW after the DR.

\begin{figure}
    \centering
    \includegraphics[width=0.9\hsize]{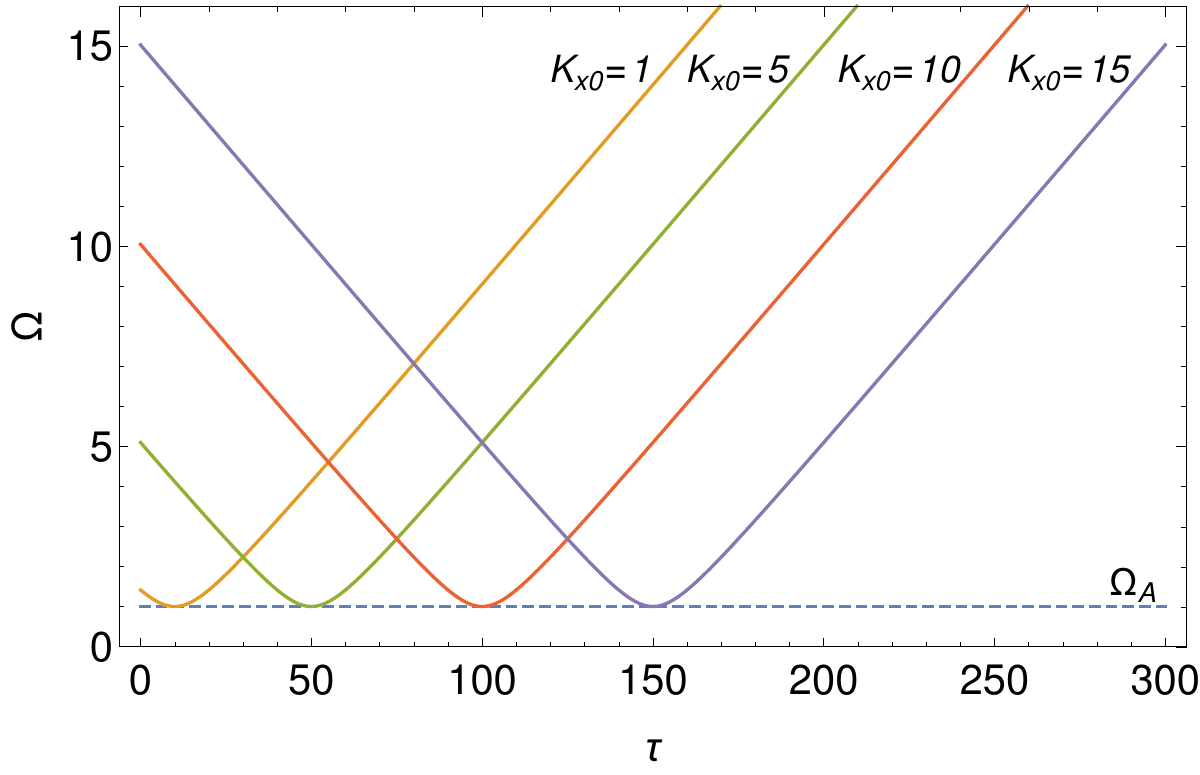}
    \caption{Dispersion curves (wave frequencies vs time) of the FMW (solid lines) and the AW (dashed line) for $R=0.1$, $K_y=0.1$, and various values of $K_{x0}$ indicated next to each line. The AW frequency is the same in all cases.}
    \label{fig:Ideal_case_freq_Kx0_variation}
\end{figure}

\begin{figure}
    \centering
    \includegraphics[width=0.9\hsize]{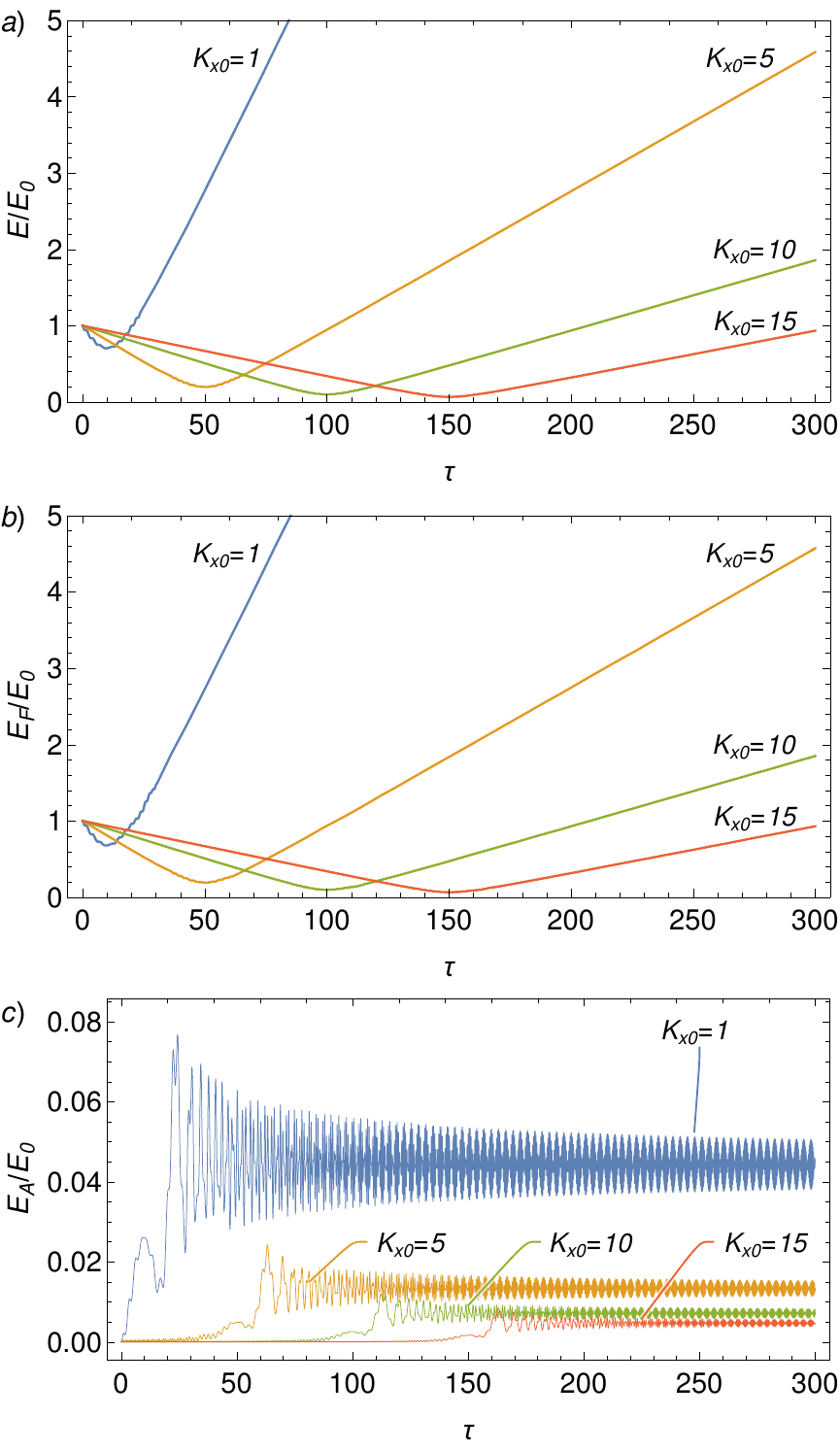}
    \caption{Ideal FMW-AW transformation with $R=0.1$,  $K_y=0.1$, and various values of $K_{x0}$ indicated within the panels. Temporal evolution of a) normalized total energy, $E(\tau)/E(0)$, b) normalized FMW energy contribution, $E_F(\tau)/E(0)$, and c) nomalized AW energy contribution, $E_A(\tau)/E(0)$.}
    \label{fig:Ideal_case_E_Kx0_variation}
\end{figure}

As shown in Figure \ref{fig:Ideal_case_freq_Kx0_variation}, the value of $K_{x0}$ determines the initial FMW frequency (Equation~(\ref{eqn:sys_freqs_TRS_FMW_AW_ideal}a)). Varying $K_{x0}$ produces a shift in the temporal evolution of the frequencies, modifying linearly the time required for the waves to reach the DR according to Equation~(\ref{eq:taustar}). However, the rate at which the FMW frequency evolves due to the flow remains constant, although the rate of energy exchange varies for each case.  This can be seen in Figure~\ref{fig:Ideal_case_E_Kx0_variation}, which displays the temporal dependence of the wave energies. Due to the different slopes that the FMW energy curves have for the various values of $K_{x0}$, and the different amplitudes of the resulting AW energies, we find that the smaller $K_{x0}$, the higher the efficiency of energy transmission from the FMW to the AW, and the earlier this coupling happens.

To understand this result, let us compare the total energy evolution (Figure~\ref{fig:Ideal_case_E_Kx0_variation}a), with the respective frequency evolution (Figure~\ref{fig:Ideal_case_freq_Kx0_variation}). It becomes evident that the relation $E\propto\Omega$ holds true outside the DR, which is consistent with the expected behavior of slowly evolving wave modes. Furthermore, its clear that the value of $K_{x0}$ modifies the proportionality constant between the energy and frequency, resulting in smaller slopes in the energy evolution for larger $K_{x0}$. The case with $K_{x0}=1$ must be highlighted, since it is already located within the DR at the beginning of the evolution, but maintains the relation $E\propto\Omega$ outside of it. The difference in $\tau_*$ of the cases with different $K_{x0}$ implies that systems with larger $K_{x0}$ take longer to reach the DR. This means that the larger the value of $K_{x0}$, more time the FMW has to give energy to the flow, and the lower the FMW energy is when the interaction with the AW begins, as can be observed in Figure \ref{fig:Ideal_case_E_Kx0_variation}a), resulting in a smaller amount of energy transferred to the AW. Hence,  the larger $K_{x0}$, the less energy is transmitted to the AW, as we can see in Figure~\ref{fig:Ideal_case_E_Kx0_variation}c).

\begin{figure}
    \centering
    \includegraphics[width=0.85\hsize]{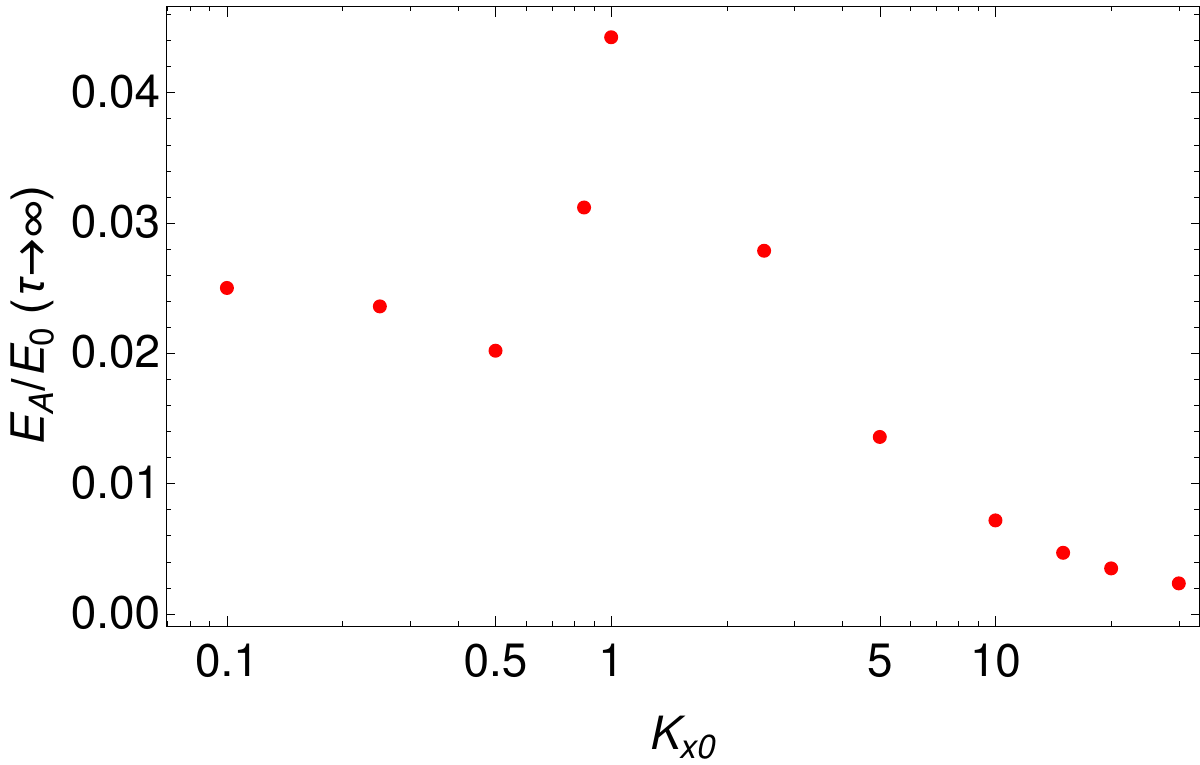}
    \caption{Fraction of energy transmitted to the AW as a function of $K_{x0}$ for an ideal FMW-AW transformation with $R=0.1$ and  $K_y=0.1$.}
    \label{fig:Ideal_Case_AW_E_tendency_regression_vars_Kx0}
\end{figure}

Figure \ref{fig:Ideal_Case_AW_E_tendency_regression_vars_Kx0} shows the  fraction of energy transmitted to the AW as a function of $K_{x0}$. The efficiency of the energy transmission  primarily follows an overall decreasing tendency with increasing $K_{x0}$, except in the proximity of $K_{x0} \approx 1$ where the AW energy fraction reaches a maximum of $5\%$, approximately. We recall that the case with $K_{x0}=1$ corresponds to a system where the DR is initially realized. Surprisingly, we find an energy transference of  $\sim 2.5\%$ when $K_{x0}\to 0$. This behavior may be unexpected, as one might not anticipate any energy transference at all when $K_{x0}=0$. However, due to the effect produced by the shear flow, the effective  wavenumber in the $x$ direction, $K_x(\tau)=K_{x0}-R\tau$, does not remain null throughout the temporal evolution and there is always  some partial wave transformation when $K_{x}\neq 0$.

\subsection{Dependence on $K_{y}$} \label{Vars Ky}

Here we study the effect of $K_y$. The value of this wavenumber component is very important for the wave coupling even in the absence of flow. \citet{2004GogoberidzeLinearCoupling} derived some analytical approximations of transformation coefficient for the FMW-AW coupling, $T_{FA}$, that informs of how much energy is transmitted between modes as the system traverses the DR. The approximations were done in the low-$\beta$ regime and as functions of the parameter $\delta=K_y/R^{1/3}$. When $\delta\gg1$, the approximated transformation coefficient between FMW and AW is
\begin{equation}
    T_{FA}\approx\frac{\pi}{2}\exp{\left(-\frac{K_y^3}{3R}\right)},
    \label{eqn:TFA_high_delta}
\end{equation}
while when  $\delta\ll1$ the  transformation coefficient can be approximated by
\begin{equation}
    T_{FA}\approx\frac{2^{2/3}\pi K_y}{3^{1/3}\Gamma\left(\frac{1}{3}\right)R^{1/3}}\left(1-\frac{\Gamma\left(\frac{1}{3}\right)}{2^{7/4}3^{1/3}\Gamma\left(\frac{2}{3}\right)}\frac{K_y^4}{R^{4/3}}\right),
    \label{eqn:TFA_low_delta}
\end{equation}
 where $\Gamma$ is the so-called gamma function. Expressions (\ref{eqn:TFA_high_delta}) and (\ref{eqn:TFA_low_delta})  fail when $\delta \sim 1$ and so $K_y\sim R^{1/3}$. Despite this, Equation (\ref{eqn:TFA_low_delta}) may be used to roughly calculate the critical $K_y$ for which  the energy transmission is maximal, namely
\begin{equation}
    K_{ycr}=\left(\frac{2^{7/4}3^{1/3}\Gamma\left(\frac{2}{3}\right)}{5\Gamma\left(\frac{1}{3}\right)}\right)^{1/4}R^{1/3},
    \label{eqn:critical_ky}
\end{equation}
\noindent which is used as a reference in this study. We recall that Equations (\ref{eqn:TFA_high_delta})--(\ref{eqn:critical_ky}) are taken from \citet{2004GogoberidzeLinearCoupling}, where all the details about their derivation can be found.

With all of this in mind, we now consider a system with $K_{x0}=10$ and $R=0.1$. According to Equation~(\ref{eqn:critical_ky}), the critical $K_y$ is  $K_{ycr}\approx 0.39$. We compute numerical solutions of this system for differnt values of $K_y$.  We also consider the case with $K_y=0$ for which the FMW and the AW become uncoupled, resulting only in the interaction between the FMW and the background flow. The variation of the initial FMW frequency with $K_y$ is negligible, since $K_{x0} \gg K_y$ in all considered cases. In addition, all cases keep the same critical time for the DR realization, $\tau_*=100$, and the same initial energy. 

\begin{figure*}
    \centering
    \includegraphics[width=0.9\hsize]{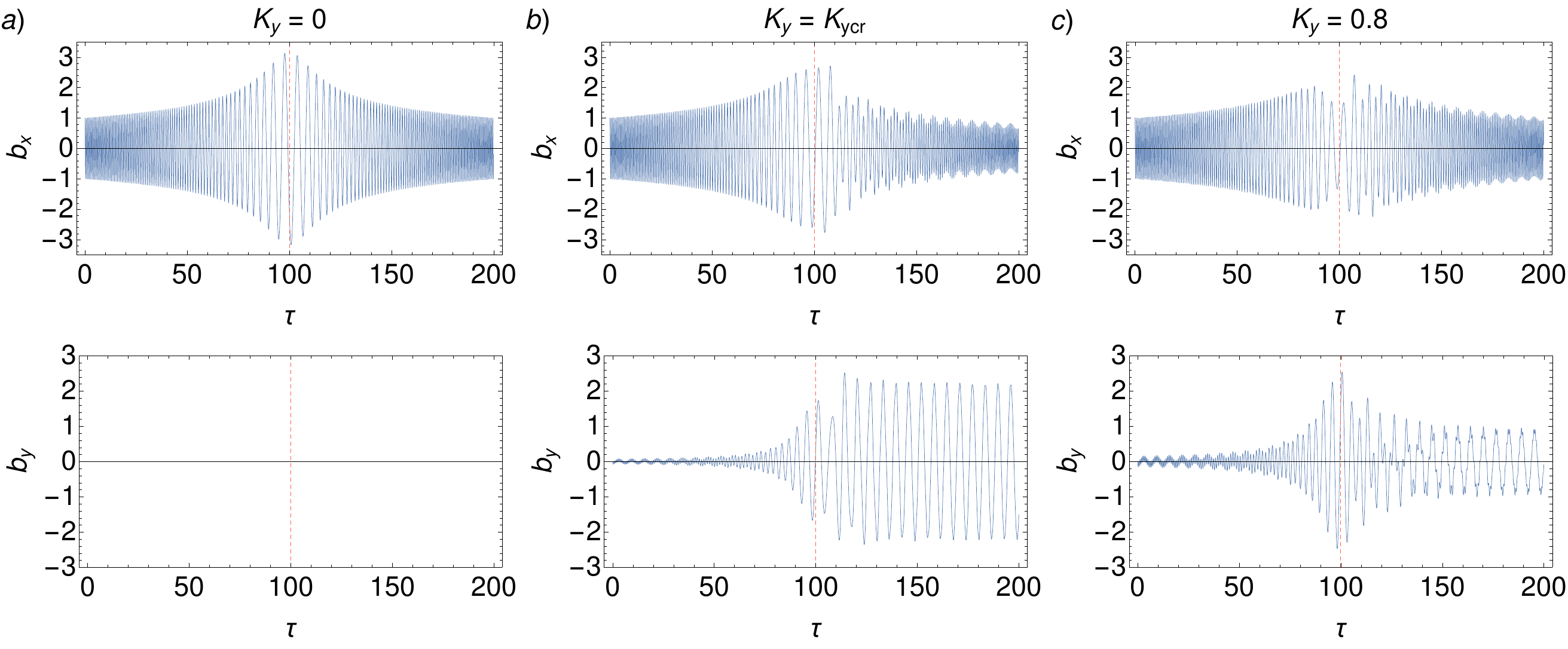}
    \caption{Ideal FMW-AW transformation with $R=0.1$ and $K_{x0}=10$ for a) $K_{y}=0$, b) $K_y = K_{ycr} = 0.39$, and c) $K_y = 0.8$. Top and bottom panels display the temporal evolutions of $b_x(\tau)$ and $b_y(\tau)$,  respectively. The red dashed line marks $\tau_*$ in all plots.}
    \label{fig:Ideal_case_bx_by_Ky_variation}
\end{figure*}

\begin{figure}
    \centering
    \includegraphics[width=0.9\hsize]{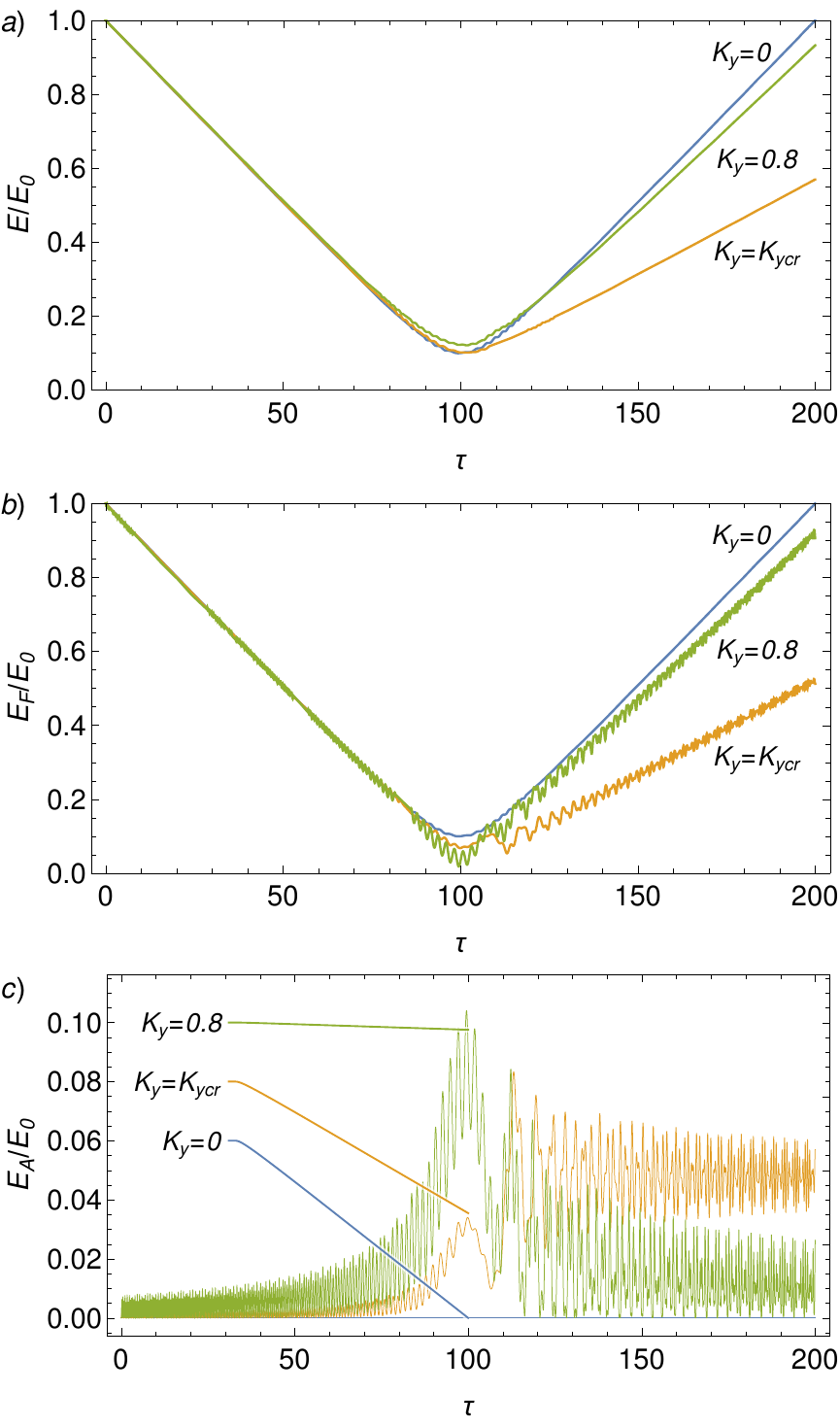}
    \caption{Same as Figure~\ref{fig:Ideal_case_E_Kx0_variation} but  with $R=0.1$,  $K_{x0}=10$, and various values of $K_{y}$ indicated within the panels.}
    \label{fig:Ideal_case_E_Ky_variation}
\end{figure}

Figure~\ref{fig:Ideal_case_bx_by_Ky_variation} shows the temporal evolution of $b_x$ and $b_y$ and Figure~\ref{fig:Ideal_case_E_Ky_variation} the corresponding energy evolution for $K_y=0$, $K_y = K_{ycr} = 0.39$, and $K_y = 0.8$. As seen in Figure~\ref{fig:Ideal_case_E_Ky_variation}, when $K_y=0$, no energy is transmitted to the AW, and the FWM just experiences the evolution caused by the shear flow: it transfers energy to the background flow until $\tau=\tau_*$, when the flow produces its reflection, and  extracts energy back from the flow afterwards. Since there is no energy interchanged with the AW in this case, the rate of energy transfer between the flow and the FMW is the same before and after the DR, but the energy flux is reversed, meaning that the FMW regains its initial energy at a time $\tau=2\tau_*$.

When we consider $K_y \neq 0$ but still $K_y\ll1$, the effects of the coupling between waves begin to appear, but they are  weak, resulting only in small scale perturbations in the Alfv\'en components. When larger values of  $K_y$  are considered, the coupling becomes more important, increasing the energy transferred to the AW. As  $K_y$ approaches $K_{ycr}$, where the maximum energy transfer between waves occurs, the perturbations begin to display mixed wave behaviors from $\tau=\tau_*$ onwards. Since a larger fraction of the energy goes to the AW, which cannot interact with the flow, there is a decrease in the efficiency at which the FMW extracts energy from the flow for $\tau>\tau_*$, as we can see from Figure \ref{fig:Ideal_case_E_Ky_variation} a, and b. As $K_y$ keeps increasing from $K_{ycr}$, the FMW and the AW begin to decouple, as the value of $K_y$ now has a non-negligible contribution to the FMW frequency at $\tau=\tau_*$. Since the FMW frequency and the AW frequency are no longer equal at $\tau=\tau_*$, a true DR cannot be realized. As a  consequence of that, the energy transmitted to the AW decreases as $K_y$ further increases from $K_{ycr}$.

For $K_y < K_{ycr}$, the energy feedback from the AW to the FMW is irrelevant, meaning that all the energy transmitted to the AW is retained in that oscillation mode. This is no longer true for $K_y \gtrsim K_{ycr}$. In the temporal evolution of the AW energy (Figure~\ref{fig:Ideal_case_E_Ky_variation}c), we observe a pulse-like phenomenon around $\tau=\tau_*$, before the system reaches its final energy state. For small $K_y$, the pulse has a very small amplitude and minimal impact on wave evolution. However, as $K_y$ increases, the amplitude of the pulse grows and, consequently, the feedback of the AW on the FMW becomes more significant. This is evident in the emergence of some mixed wave behaviors in the perturbations (Figure~\ref{fig:Ideal_case_bx_by_Ky_variation}).

When $K_y$ takes large values, another feature begins to appear in the temporal evolution of the waves. It is evident in  Figure~\ref{fig:Ideal_case_bx_by_Ky_variation} that for $K_y = 0.8$ the $b_x$ amplification around $\tau=\tau_*$ becomes truncated by a sudden amplitude decrease. This phenomenon is accompanied by an increase in the amplitude of $b_y$, during which the $b_y$ oscillation has the frequency of the FMW rather than that of the AW. This coincides with a maximum in the energy transmitted to the AW, after which the AW quickly returns a fraction of its energy back to the FMW and recovers its characteristic AW frequency. This feature of the transformation becomes more pronounced as the value of $K_y$ rises.

\begin{figure}
    \centering
    \includegraphics[width=0.85\hsize]{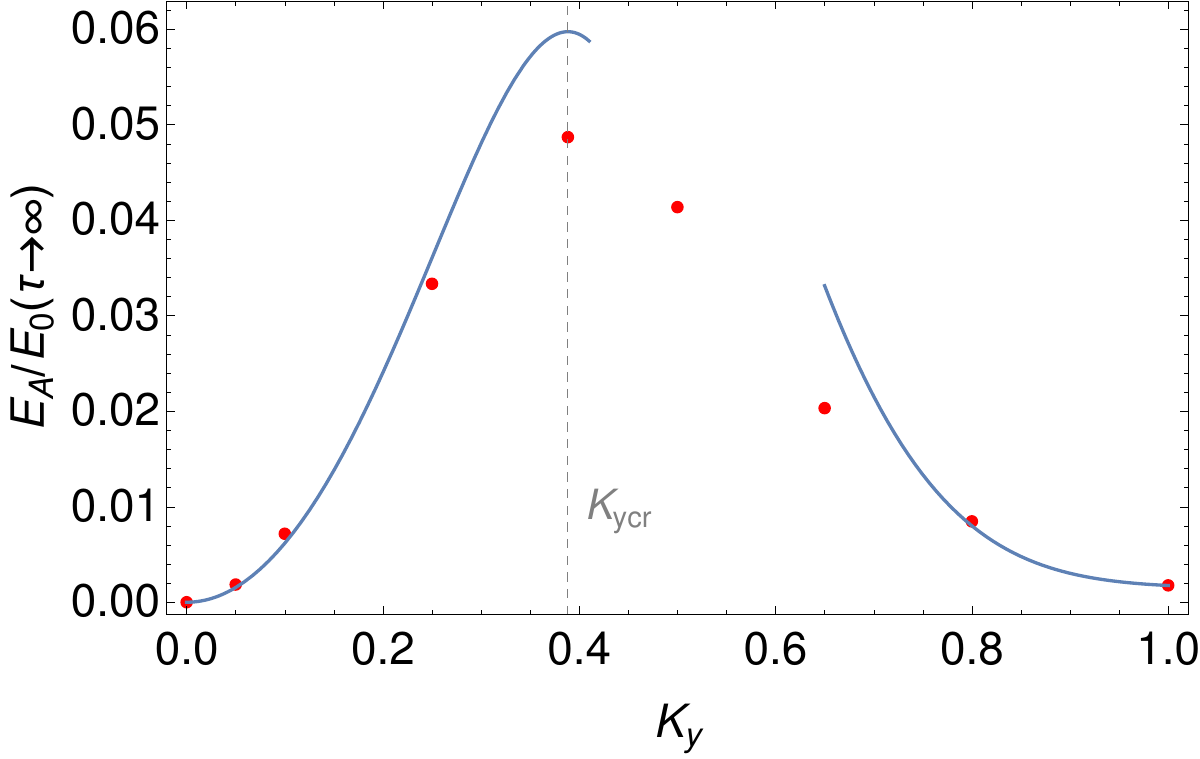}
    \caption{Fraction of energy transmitted to the AW as a function of $K_{y}$ for an ideal FMW-AW transformation with $R=0.1$ and  $K_{x0}=10$. The red points correspond to the numerical results, while the blue lines are the approximated transmission coefficients from \citet{2004GogoberidzeLinearCoupling}. The vertical dashed line denotes $K_{ycr}$.}
    \label{fig:Ideal_Case_AW_E_tendency_regression_vars_Ky}
\end{figure}

 An analysis of the  fraction of energy transmitted to the AW as a function of $K_{y}$ has been conducted. In Figure~\ref{fig:Ideal_Case_AW_E_tendency_regression_vars_Ky}, the results obtained from the numerical calculations are compared to the transformation coefficients of \citet{2004GogoberidzeLinearCoupling}, given in Equations~(\ref{eqn:TFA_high_delta}) and (\ref{eqn:TFA_low_delta}). The square of the analytical coefficients are plotted, since they are transformation coefficients for the wave amplitude and the energy depends quadratically on the amplitudes. Moreover, the coefficients need to be scaled by multiplying them by a constant to correctly match the numerical data. It is clear from Figure~\ref{fig:Ideal_Case_AW_E_tendency_regression_vars_Ky} that the analytical coefficients fit the tendency of the numerical results reasonably well in their corresponding range of applicability, but significant discrepancies appear when $K_y\approx K_{ycr}$. Nevertheless, the utility of the analytical approximations of the transmission coefficients derived by \citet{2004GogoberidzeLinearCoupling} is evident. As expected, the  energy transmission to the AW is maximal when $K_y = K_{ycr}$ and is $\sim 5\%$ the initial wave energy.

\subsection{Dependence on $R$} \label{Vars R}

\begin{figure*}
    \centering
    \includegraphics[width=0.9\hsize]{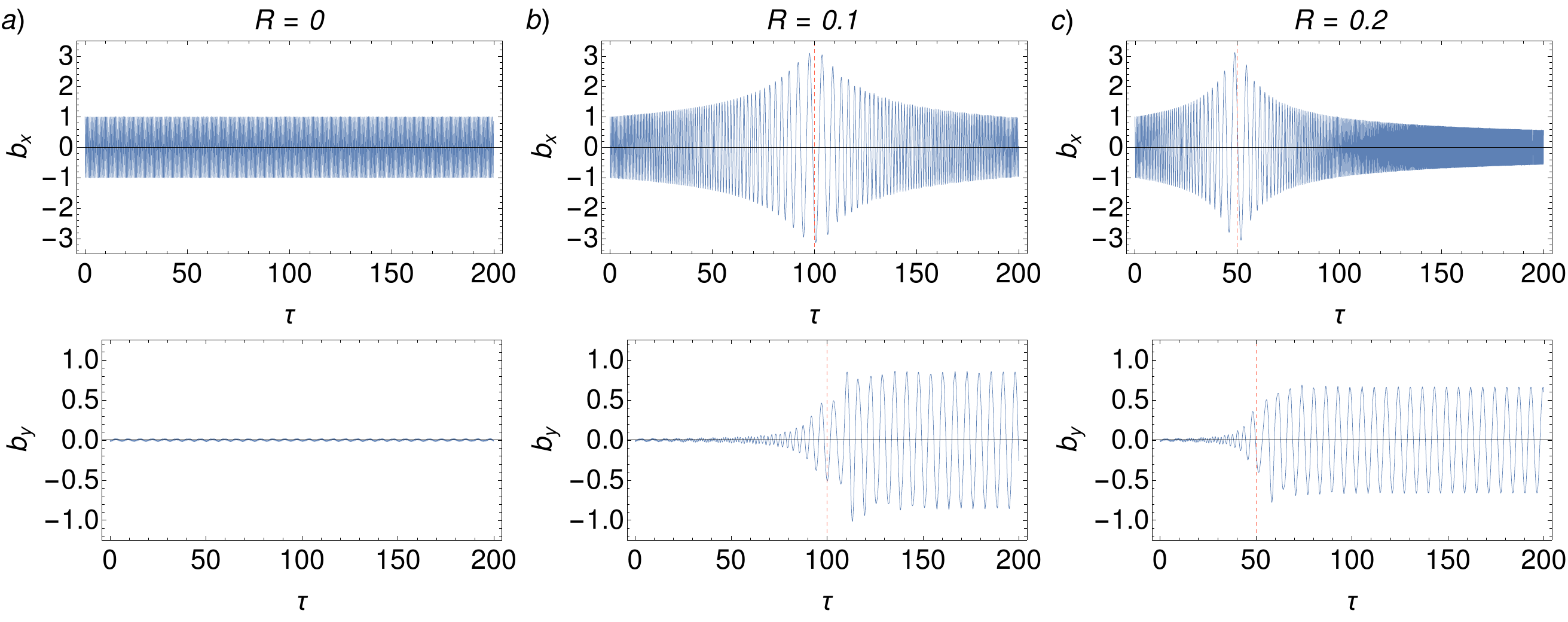}
    \caption{Ideal FMW-AW transformation with $K_{x0}=10$ and $K_y=0.1$ for a) $R=0$, b) $R = 0.1$, and c) $R = 0.2$.  Top and bottom panels display the temporal evolutions of $b_x(\tau)$ and $b_y(\tau)$,  respectively. The red dashed line marks $\tau_*$ in all plots.}
    \label{fig:Ideal_case_bx_by_R_variation}
\end{figure*}

\begin{figure}
    \centering
    \includegraphics[width=0.9\hsize]{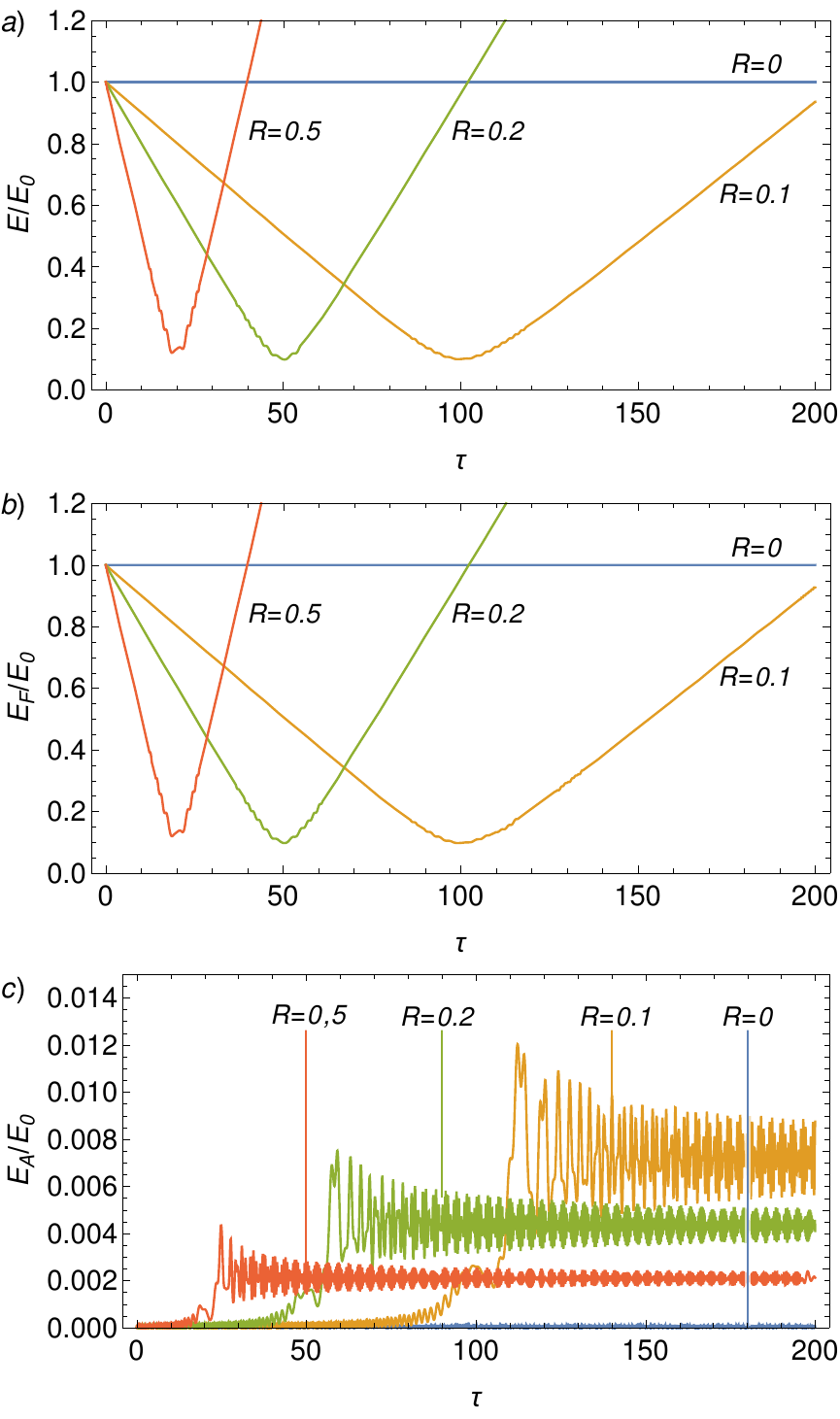}
    \caption{ Same as Figure~\ref{fig:Ideal_case_E_Kx0_variation} but with $K_{x0}=10$,  $K_y=0.1$, and various values of $R$ indicated within the panels.}
    \label{fig:Ideal_case_E_R_variation}
\end{figure}

Finally in this ideal case, we study the dependence on the normalized strength of the velocity shear, $R$.
We consider  $K_{x0}=10$ and $K_y=0.1$, while $R$ is varied. Figures~\ref{fig:Ideal_case_bx_by_R_variation} and \ref{fig:Ideal_case_E_R_variation} display the evolution of  the perturbations ($b_x$ and $b_y$) and the wave energies, respectively, for $R=0$, 0.1, and 0.2, with the energy plot also showing an additional result for $R=0.5$. The shearless scenario with $R=0$ is  included as a reference, as  there are no effects due to the  flow and no energy exchange  between waves. The excitation of the AW is minimal in the shearless case and is only caused by the fact that $K_y\neq0$.

\begin{figure}
    \centering
    \includegraphics[width=0.9\hsize]{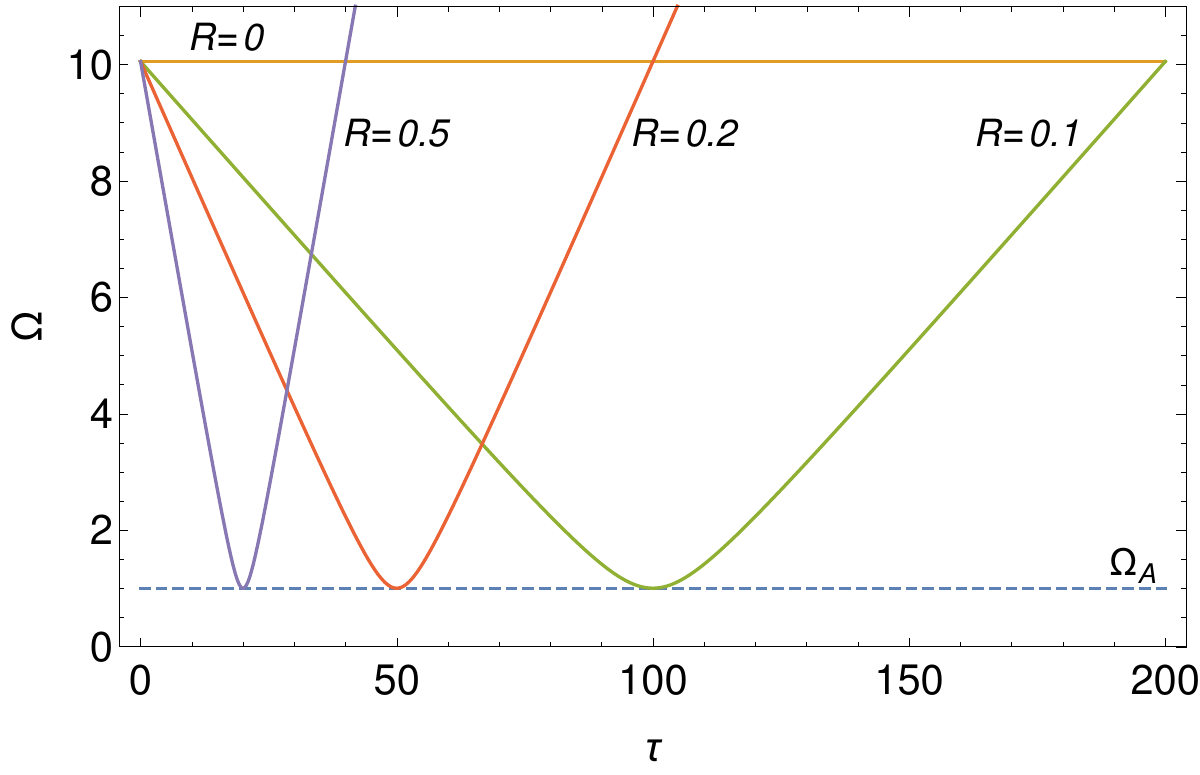}
    \caption{Same as Figure~\ref{fig:Ideal_case_freq_Kx0_variation} but with $K_{x0}=10$, $K_y=0.1$, and various values of $R$ indicated within the figure.}
    \label{fig:Ideal_case_freq_R_variation}
\end{figure}

When $R\neq 0 $, there is partial transformation of the FMW into the AW. The time for the DR realization, $\tau=\tau_*$, depends on $R$ accoding to Equation~(\ref{eq:taustar}). Hence, the larger $R$, the smaller $\tau_*$. In contrast with what happens when $K_{x0}$ is varied, where a temporal shift of the whole evolution is found, now there is no change in the initial FMW frequency, when $R$ changes, and the time scale of energy exchange between background flow and the FMW is affected. This can be better seen by comparing Figures~\ref{fig:Ideal_case_freq_Kx0_variation} and \ref{fig:Ideal_case_freq_R_variation}. In the latter it is seen that the larger $R$, the steeper the frequency slope. This agrees with the discussion in Section \ref{chap:freqs-E} where it is noted that the time scale of the energy exchange between the waves and the flow is of the order of $\sim R^{-1}$.

\begin{figure}
    \centering
    \includegraphics[width=0.85\hsize]{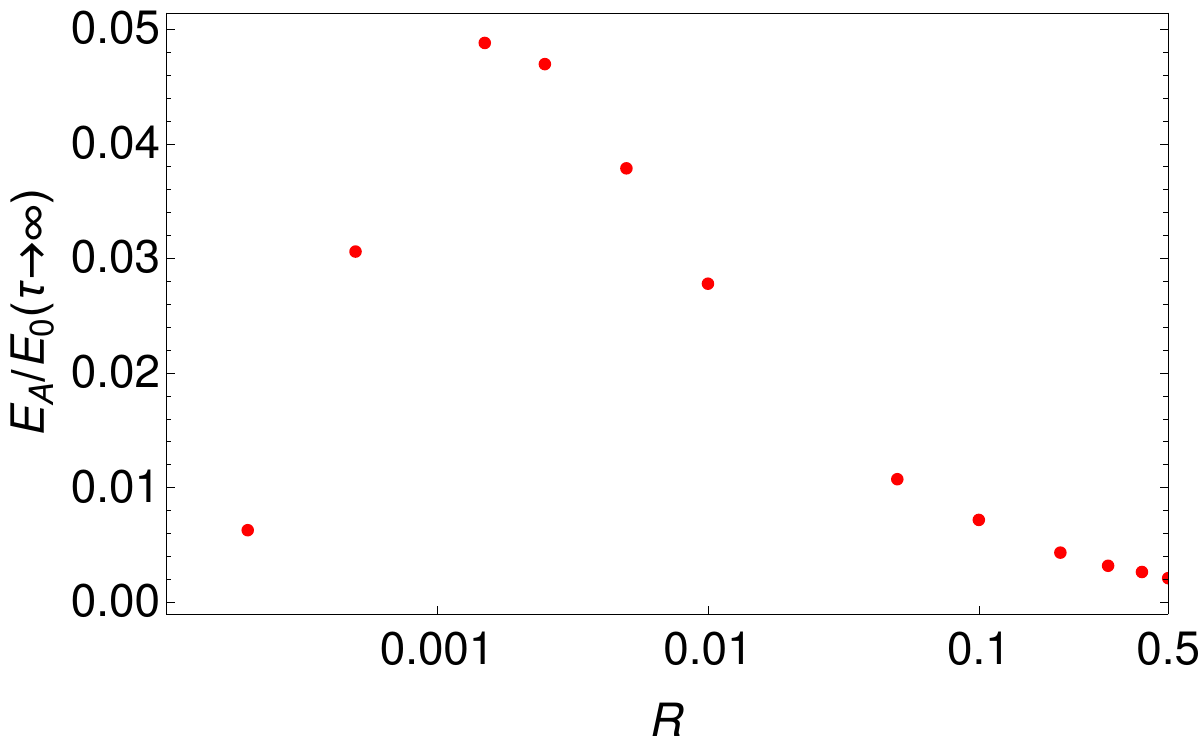}
    \caption{Fraction of energy transmitted to the AW as a function of $R$ for an ideal FMW-AW transformation with  $K_{x0}=10$ and $K_y = 0.1$.}
    \label{fig:Ideal_Case_AW_E_tendency_regression_vars_R}
\end{figure}

The time scale of interaction between waves within the DR also changes with the value of $R$, since the rate of change of $K_x(\tau)$ is proportional to $R$. Specifically, the larger  $R$, the shorter the time spent in the DR. This may conflict with the condition of ``slow passing" through the DR if $R$ is large enough, thereby reducing the efficiency of energy transfer between waves. This behavior is confirmed in  Figure \ref{fig:Ideal_case_E_R_variation}c), where the  energy transmitted to the AW decreases as the value of $R$ increases, and it becomes even clearer in the energy analysis of Figure~\ref{fig:Ideal_Case_AW_E_tendency_regression_vars_R}. We found that the energy transference grows as $R$ increases from $R=0$ until a maximum is reached around $R\approx 1.5\times10^{-3}$, where the energy transfer to the AW is $5\%$ of the initial energy, approximately. From this point,  the energy transferred to the AW begins to decrease rapidly as $R$ keeps increasing. The  occurrence of the maximum is caused by the competition of two already discussed temporal scales, namely the time of energy exchange between the flow and the FMW (that decreases with increasing $R$) and the time spent in the DR.

Lastly, special attention has to be paid to the case with $R=0.5$, which may not meet the condition of low shear. However, the results show that the energy follows closely the relation $E\propto\Omega$ when the system is outside the DR. Additionally, the behavior of the perturbations $b_x$ and $b_y$ (not shown here) display a consistent behavior with the rest of the studied cases. This suggests that the approximation $E\propto\Omega$ can be applied successfully over a wider range of $R$ values beyond the strict low shear limit.

\section{FMW-AW transformation with ambipolar diffusion} \label{Vars eta}

Now that we have a clear picture on how the mode coupling process behaves in the ideal case, we introduce the effect of ambipolar diffusion, so that $\tilde{\eta}_A\neq0$. We retain $\varepsilon=0$ to  study the FMW-AW transformation alone. These conditions reduce Equations~(\ref{eqn:sys_eqs_11_diffusion}) to 
\begin{subequations}
    \begin{align}
        \begin{split}
            \label{eqn:equation_11a_diffusion_simplif}
            \psi^{(2)} -\tilde{\eta}_A R \left(2+4K_x^2+K_y^2\right)b_x-3\tilde{\eta}_A R K_xK_yb_y&+\\
            +\tilde{\eta}_A K_x\left(1+K_x^2+K_y^2\right)b_x^{(1)}+\tilde{\eta}_A K_y\left(1+K_x^2+K_y^2\right)b_y^{(1)}&=0,
        \end{split}
        \\
        \begin{split}
            \label{eqn:equation_11b_diffusion_simplif}
            b_x^{(2)}+\tilde{\eta}_A\left(1+K_x^2\right)b_x^{(1)}+\left(1+K_x^2-2\tilde{\eta}_A RK_x\right)b_x &+\\
            +\tilde{\eta}_A K_xK_yb_y^{(1)}+\left(K_xK_y-\tilde{\eta}_A RK_y\right)b_y &= 0,
        \end{split}
        \\
        \begin{split}
            \label{eqn:equation_11c_diffusion_simplif}
            b_y^{(2)} + \tilde{\eta}_A\left(1+K_y^2\right)b_y^{(1)} + \left(1+K_y^2\right)b_y &+\\
            +\tilde{\eta}_A K_xK_yb_x^{(1)} + \left(K_xK_y-\tilde{\eta}_A RK_x\right)b_x &= 0,
        \end{split}
    \end{align}
    \label{eqn:sys_Trs_FMW_AW_diffusion}
\end{subequations}
which describe the interaction between the FMW and the AW, but now adding the ambipolar diffusion. Contrary to what happens in the  ideal System~(\ref{eqn:sys_Trs_FMW_AW_ideal}), the equation for $\psi$ (Equation~(\ref{eqn:equation_11a_diffusion_simplif})) cannot be discarded, as it becomes coupled with $b_x$ and $b_y$ through the  ambipolar diffusion terms. However, $b_x$ and $b_y$ remain uncoupled to $\psi$ because in Equations~(\ref{eqn:equation_11b_diffusion_simplif}) and (\ref{eqn:equation_11c_diffusion_simplif}) there are  no terms with $\psi$. Therefore, the evolution of $\psi$  affects neither the FMW nor the AW, which means that for some of the calculations it is enough to consider the latter two equations alone. Conversely, the waves would drive perturbations in $\psi$, i.e., perturbations in density, which can be computed afterwards from the first equation once $b_x$ and $b_y$ are known.

As the diffusion terms are presumably small for physical conditions in the partially ionized solar plasma \citep[see, e.g.,][]{Khodachenko2004, 2024SolerMHDWavesinPIP}, we may still use the fundamental frequencies for the ideal case (Equations~(\ref{eqn:sys_freqs_TRS_FMW_AW_ideal})) as approximations for the frequencies of the waves in the presence of dissipation. Furthermore, the energy of the waves can still be computed with the expressions in Equation (\ref{eqn:Energy_F_A}), as the change in the amplitude of the perturbations themselves account for the diffusion-related damping.

The set-up we analyze here is equivalent to the one  studied in the ideal scenario of an initially pure FMW that partially transforms to an AW, obtained using the initial conditions $b_x(0)=1$, and $b_y(0)=b_x'(0)=b_y'(0)=0$. Concerning the conditions for $\psi$, we use $\psi(0)=K_{x0}b_x(0)$ and $\psi'(0)=0$, thus ensuring that the initial  density perturbation is zero. As system parameters we take $K_{x0}=10$, $K_y=0.1$ and $R=0.1$, which result in $\tau_*=100$.
The dimensionless diffusion coefficient, $\tilde{\eta}_A$, is taken as a free parameter.

\begin{figure*}
    \centering
    \includegraphics[width=0.9\hsize]{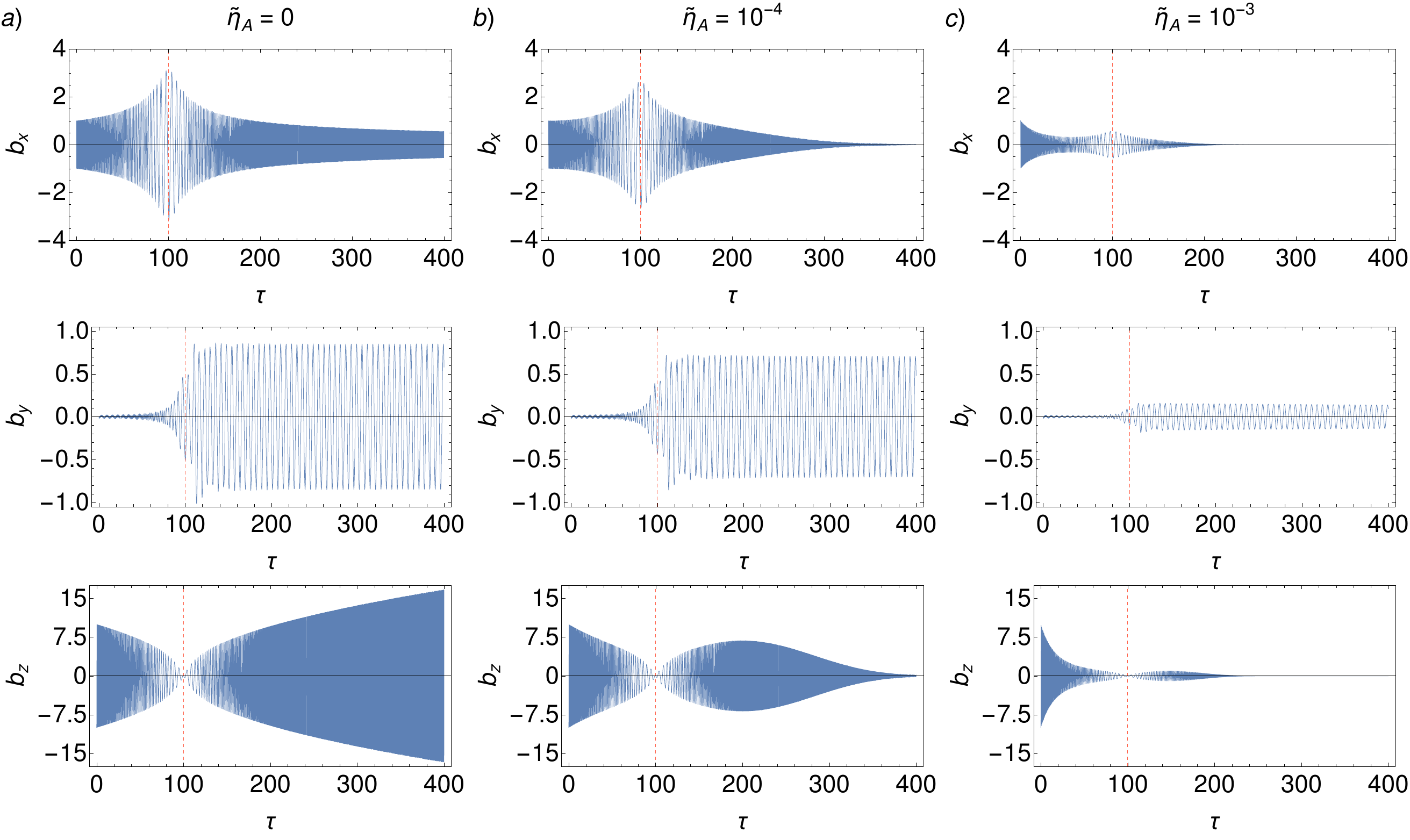}
    \caption{Dissipative FMW-AW transformation with $R=0.1$, $K_{x0}=10$, and $K_y=0.1$ for a) $\tilde{\eta}_A=0$, b) $\tilde{\eta}_A=10^{-4}$, and c) $\tilde{\eta}_A=10^{-3}$. From top to bottom it is displayed the temporal evolutions of $b_x(\tau)$,  $b_y(\tau)$,  and $b_z(\tau)$,   respectively. The red dashed line marks $\tau_*$ in all plots.}
    \label{fig:diffusion_case_bx_by_bz_eta_variation}
\end{figure*}

\begin{figure}
    \centering
    \includegraphics[width=0.9\hsize]{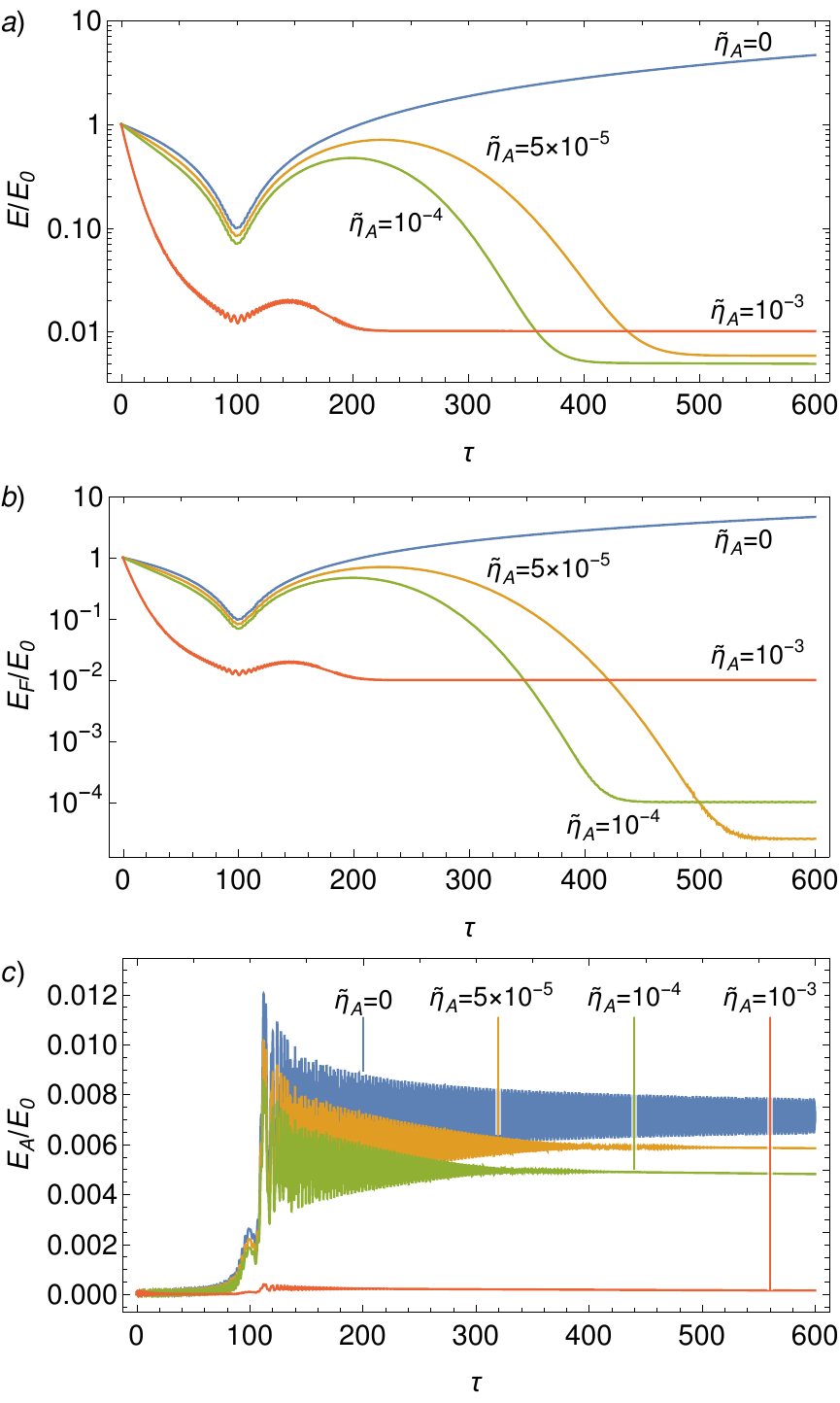}
    \caption{Same as Figure~\ref{fig:Ideal_case_E_Kx0_variation} but for a dissipative FMW-AW transformation with $R=0.1$,  $K_{x0}=10$, $K_y=0.1$ and various values of $\tilde{\eta}_A$ indicated within the panels.}
    \label{fig:diffusion_case_E_eta_variation}
\end{figure}

Figure~\ref{fig:diffusion_case_bx_by_bz_eta_variation} shows the evolution of the magnetic field perturbations $b_x$, $b_y$, and $b_z$ for $\tilde{\eta}_A = 10^{-4}$ and $\tilde{\eta}_A = 10^{-3}$. Additionally, the solutions in the ideal case (obtained with $\tilde{\eta}_A=0$) are shown as a reference. The first effect we observe is the expected decrease of the waves amplitude for large times when ambipolar diffusion is present. The waves exhibit a faster damping as $\tilde{\eta}_A$ increases. Valuable information is provided by the analysis of the energy evolution in Figure~\ref{fig:diffusion_case_E_eta_variation}. The observed damping of the perturbations is accompanied by a dissipation of wave energy. A consequence of the energy dissipation is a change of the behavior of the energy temporal evolution since, as commented earlier, in the last lines of Subsect.~\ref{chap:freqs-E}, we now have $E\not\propto\Omega$.
The difference between the FMW energy evolution in the ideal case and that with diffusion becomes substantial at large times, and the larger $\tilde{\eta}_A$, the earlier this difference arises. Another property of the total energy evolution is that it saturates to a constant value for large times and this saturation value gets larger as $\tilde{\eta}_A$ increases. This feature of the energy evolution is caused by a new effect that is absent from the the ideal scenario and appears when ambipolar diffusion is present. This new effect is analyzed later in the final paragraphs of this Section.

Concerning the FMW-AW conversion around $ \tau = \tau_*$, we reach the following conclusions. In the first place, because of the presence of diffusion, the FMW dissipates part of its energy before reaching the DR. This results in a reduction of the available  energy  to be transmitted to the AW. For the largest considered value of $\tilde{\eta}_A = 10^{-3}$, the energy transferred to the AW is negligible. This occurs because the FMW damping happens so fast that almost all the FMW energy has already been dissipated before reaching the DR. Secondly, in Figure~\ref{fig:diffusion_case_E_eta_variation}c) it might appear that, after the mode conversion, the AW energy stays unaffected by the diffusion as it seems to reach a constant value for large times. Detailed investigation reveals that there is indeed diffusion in the AW, but it occurs at a much slower pace compared to that of the FMW. This behavior is related to the different damping time scales that the MHD waves have in the presence of ambipolar diffusion \citep[see][]{2024SolerMHDWavesinPIP}. The AW damping rate depends only on the parallel component of the wavenumber, $k_z$, which remains constant throughout the evolution, while that of the FMW depends on the wavenumber modulus, which increases with time  owing to the effect of the shear flow on  $K_x$. At a sufficiently large time,  when the wavenumber modulus is large enough, the FMW becomes much more efficiently damped than the AW.

In the ideal case, for $\tau > \tau_*$ the shear flow provides an unlimited input of energy to the reflected FMW. Now we find that ambipolar diffusion impairs the ability of the flow to give energy to the FMW. Owing to the growing efficiency of the ambipolar damping with time, the FMW energy increase stops when the energy dissipation rate becomes equal than the energy extraction rate from the flow, and decreases afterwards. In Figure~\ref{fig:diffusion_case_E_eta_variation} it is easy to see that this maximum in the energy evolution shifts to smaller times as $\tilde{\eta}_A$ increases. A study on how the value of $\tilde{\eta}_A$ affects the maximum energy and the time needed to reach it has been carried out by considering a wide range of $\tilde{\eta}_A$. The results (not shown here for simplicity) confirm a decreasing trend in the maximum energy and the corresponding times at which the maximum is attained with increasing $\tilde{\eta}_A$.

\begin{figure}
    \centering
    \includegraphics[width=0.9\hsize]{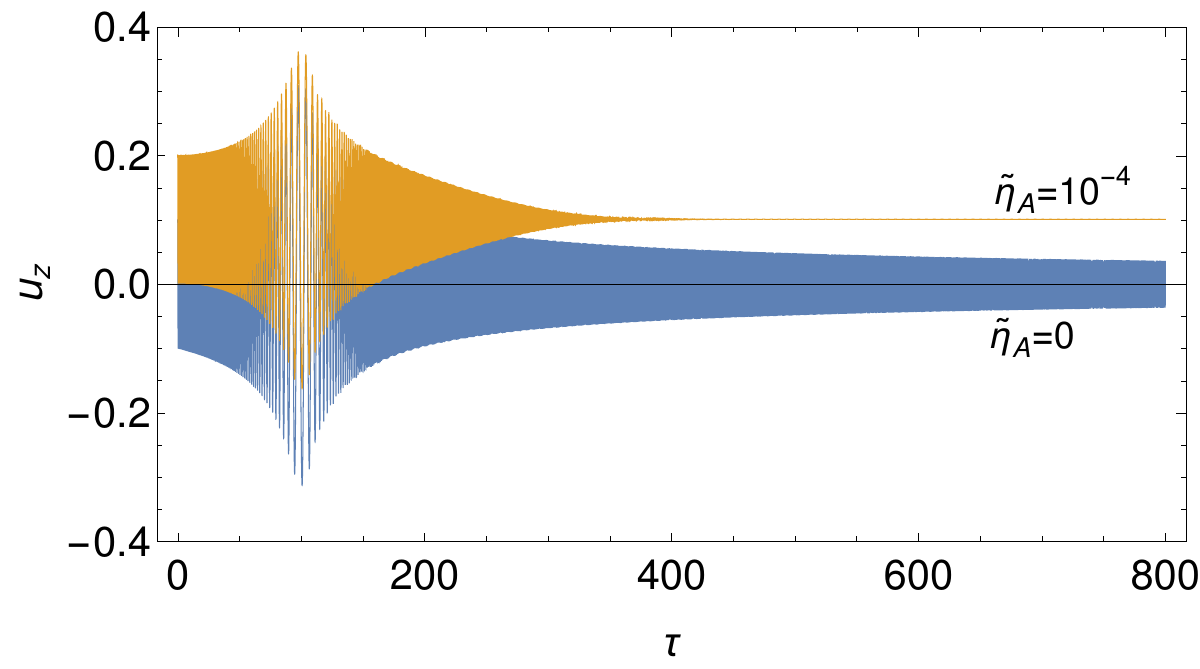}
    \caption{Temporal evolution of $u_z(\tau)$ in the ideal case (blue line) and in a dissipative case with $\tilde{\eta}_A=10^{-4}$ (orange line) for a FMW-AW transformation with $R=0.1$, $K_{x0}=10$, and $K_y=0.1$.  }
    \label{fig:diffusion_case_vz_eta_variation}
\end{figure}

As commented earlier, it is found that the wave energy saturates to a constant value at large times when ambipolar diffusion is present (see Figure~\ref{fig:diffusion_case_E_eta_variation}). This is caused by a new effect related to the  appearance of a constant flow along the $z$-direction as a consequence of the diffusion. This additional flow originates from the equation for $\psi$ (Equation (\ref{eqn:equation_11a_diffusion_simplif})), which was uncoupled in the ideal case. However,  perturbations are now driven in $\psi$ due to a diffusion-induced coupling. Perturbations in $\psi$ are associated with density perturbations in an effect akin to having a pressure gradient along the $z$-axis.
That, in turn, drive velocity perturbations in the $z$-direction. This becomes evident when examining the evolution of $u_z$ in Figure \ref{fig:diffusion_case_vz_eta_variation}, showing that the mean value of $u_z$ at  large times is zero in the ideal case but nonzero in the ambipolar case. This constant flow increases with the value of $\tilde{\eta}_A$.

To further explore the nature of this constant longitudinal flow driven by diffusion, we have extended the investigation by considering the case with gas pressure ($\varepsilon \neq 0$) and have reexamined some of the computations. These results are not included here and are to be explored in detail in the forthcoming continuation of this paper, where the case  $\varepsilon \neq 0$ will be investigated at length. However, we feel it convenient to anticipate these findings as they are crucial to understand the diffusion-induced flow found here.

When $\varepsilon \neq 0$, it turns out that this constant longitudinal flow driven by diffusion is absent from the results. Instead, the energy that would be transferred to this longitudinal flow is invested in exciting the SMW, which is  present in the system when $\varepsilon \neq 0$. Thus, it becomes clear that the true nature of this constant flow we obtain here is to produce a new coupling with the SMW mediated by ambipolar diffusion that has no equivalence in the ideal case. To the best of our knowledge, this has not been reported before in the literature. The SMW cannot be driven in the present computations with $\varepsilon=0$, but the coupling mechanism remains somehow in the form of this constant flow. This result evidences that the general $\varepsilon \neq 0$ case needs to be considered to fully understand the coupling between the MHD waves in the ambipolar case. Its comprehensive study is  beyond the scope of this initial paper and is left for the second part.

\section{Application to solar prominences} \label{app_threads}

We perform a specific application to solar prominences in order to explore the FMW-AW transformation and energy transference that may happen in these structures.  Solar prominences are large structures of relatively cool and dense plasma suspended in the solar corona. The prominence plasma  has  physical conditions similar to those found in the chromosphere \citep[see, ][]{2015VialSolarProminences}. This is partially ionized plasma with a temperature in the range of 7000--9000~K. Prominences are composed of a myriad of thin threads that outline particular field lines of their magnetic structure \citep[see, ][]{2011LinFilamentThread-likeStructures, 2014MartinMagneticFieldProminences}. It has been shown that transverse MHD waves are ubiquitous in the fine threads \citep[see, e.g.,][]{2009LinSwayingThreads,2015BallesterSolarProminences,2018ArreguiProminenceOscillations}, and there is evidence suggesting that the  waves originate at the photosphere \citep{hillier2013}. Alfv\'enic waves driven at the photosphere can transport a substantial amount of energy to  coronal heights  \citep[see][]{2019SolerEnergyTransport}

We consider a simple model of a thin thread with a length\footnote{The so-called thread length  refers only to the length of the dense plasma region, which is presumably located inside a much longer magnetic tube.} of $L_T=3\times10^{6}$~m and a radius an order of magnitude lower, $R_T=1.5\times10^{5}$~m, with a temperature of $T=8000$~K, a  density of $\rho = 5\times10^{-11}$~Kg~m$^{-3}$, and a magnetic field strength of $B_0=10$~G. The Alfv\'en speed is $V_A \approx 126$~km~s$^{-1}$. The  ambipolar diffusion coefficient of the plasma in the thread is $\eta_A=5.4\times10^7$ m$^2$~s$^{-1}$. This value has been adopted from the prominence thread models of \citet{2023MelisEquilibriumModelsofProminence}, where the details of the calculation can be found. The dimensionless ambipolar diffusion coefficient is $\tilde{\eta}_A \approx 4.5\times10^{-4}$.

The wavenumber components $k_z$ and $k_{x0}$ are related to the physical dimensions of the thread as $k_z = \pi / L_T \approx 1.05\times 10^{-6}$~m$^{-1}$ and $k_{x0} = \pi / R_T \approx 1.05\times 10^{-5}$~m$^{-1}$, resulting in a dimenssionless $K_{x0}=k_{x0}/k_z=10$.  To calculate the normalized strength of the velocity shear, $R$, we assume typical flow velocities in solar prominences \citep[see, e.g., ][]{1998ZirkerStreamingGasFlowsSolarProminences} and take a variation of the flow velocity across the thread of 10~km~s$^{-1}$, resulting in $A= 1/30$~s$^{-1}$ and thus $R \approx 0.252$. The value of $K_y$ is arbitrary and we study two situations, $K_y=0.1 K_{x0}=1$ and $K_y=K_{x0}=10$, corresponding to a kink mode and a high fluting mode, respectively, in the cylindrical equivalent to this Cartesian configuration. The value of $K_{ycr}$ according to Equation (\ref{eqn:critical_ky}) is $K_{ycr}\approx 0.53$, which is smaller than the two considered values but relatively close to the kink case. 

With all that, we have $\tau_*\approx 40$. We can reassign dimensions to this characteristic time to determine the physical time scale of the mode conversion and find  $t_*\approx 300$ s. In other words, the temporal scale of the energy exchange between background flow and wave modes is approximately 5 minutes. This is of the same order as the periods of the observed transverse oscillations in threads \citep[see, e.g., ][]{2009LinSwayingThreads}, which indicates that the shear-flow-driven wave coupling may be relevant in these structures. 

\begin{figure}
    \centering
    \includegraphics[width=0.85\hsize]{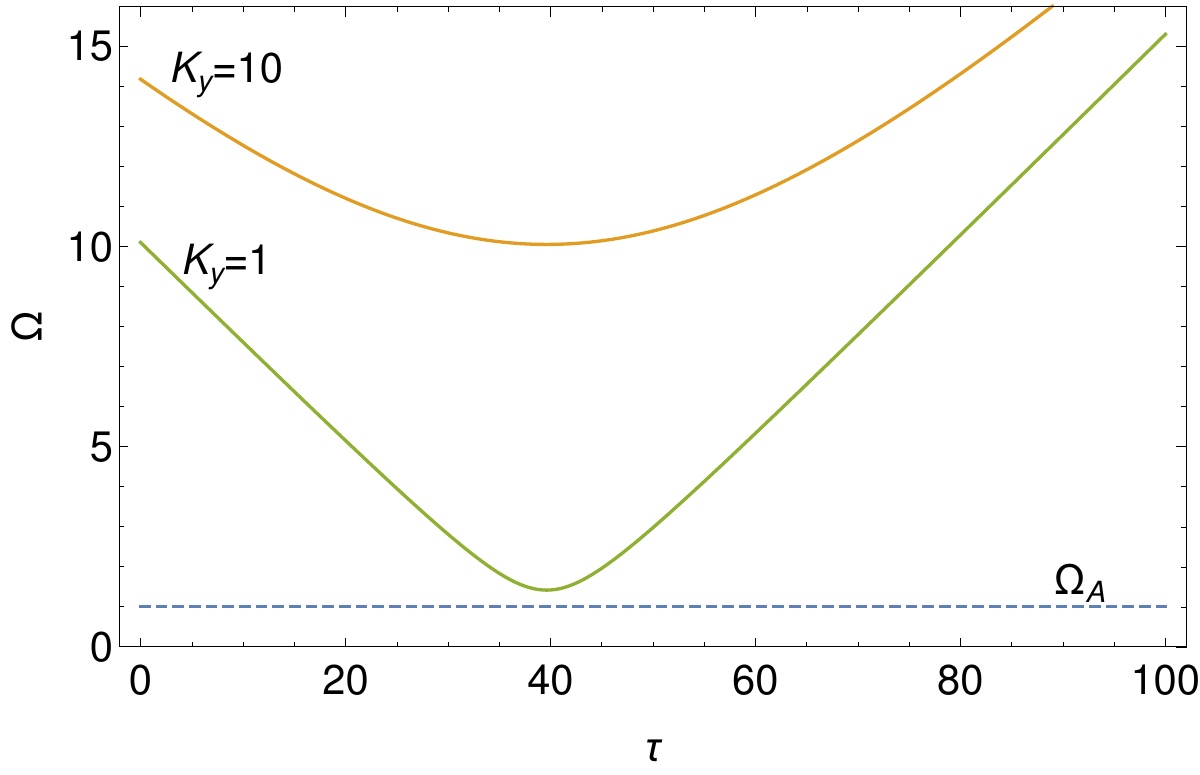}
    \caption{Dispersion curves (wave frequencies vs time) of the FMW (solid lines) and the AW (dashed line) for the parameters used in the application of Section~\ref{app_threads}.}
    \label{fig:Thread_case_Frequency}
\end{figure}

A notable aspect of this application is the significant differences between the FMW frequencies of the two cases with different $K_y$. This is evident in the dispersion curves  shown in Figure~\ref{fig:Thread_case_Frequency}. The case with $K_y=1$ presents a dispersion curve quite similar to the theoretical cases studied earlier, with a FMW frequency that becomes very close to the AW frequency when $\tau\sim\tau_*$, hence ensuring there will be some  coupling and energy transference between wave modes. Conversely, in the case with $K_y=10$ the initial FMW frequency is  higher and  never approaches the AW frequency sufficiently, which is an essential condition for the DR realization. Therefore, using again the cylindrical equivalent to this Cartesian case, we speculate that an efficient coupling may happen for kink modes but not for fluting modes in  threads modeled as cylindrical flux tubes.

\begin{figure*}
    \centering
    \includegraphics[width=0.9\hsize]{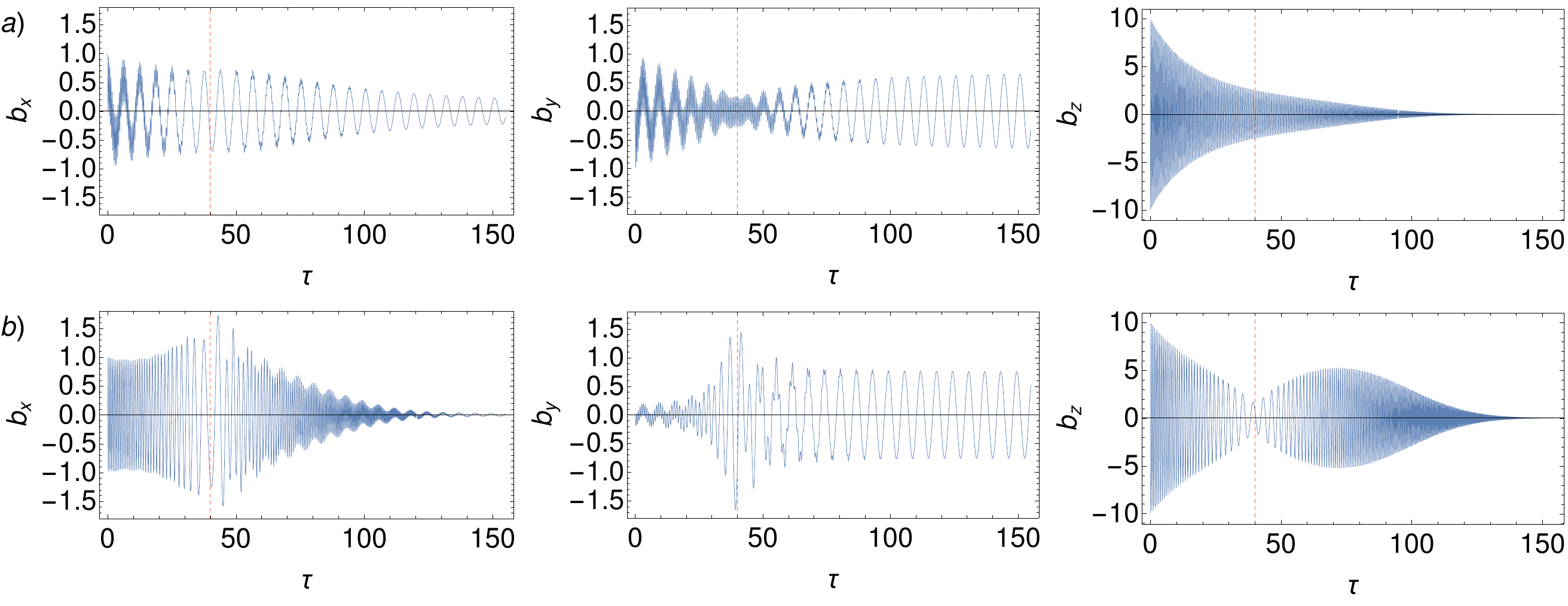}
    \caption{Dissipative FMW-AW transformation for the solar prominence thread conditions of Section~\ref{app_threads} with a) $K_y=10$ and b) $K_y=1$. From left to right it is displayed the temporal evolution of $b_x(\tau)$, $b_y(\tau)$, and $b_z(\tau)$, respectively. The red dashed line marks $\tau_*$ in all plots.}
    \label{fig:thread_case_bx_by_bz_representation}
\end{figure*}

The temporal evolution of the magnetic field perturbations are displayed in Figure~\ref{fig:thread_case_bx_by_bz_representation}. The aspect that stands out most in these results is the great difference the temporal evolution presents between both cases, as was expected. On the one hand, when $K_y=1$ it is easy to see the typical temporal evolution of an initial FMW that partially transforms into an AW around $\tau=\tau_*$, for a case with a value of $K_y$ higher than the $K_{ycr}$, but not large enough for the shear coupling not to be the primary coupling mechanism between both wave modes. We find in the $b_x$ component the initial amplification of the FMW up to $\tau\sim\tau_*$, at which point this amplification is truncated due to the momentary coupling and energy transfer between the waves that happens within the DR. This is followed  by the strong damping of the FMW by the ambipolar diffusion for $\tau>\tau_*$. The FMW is completely damped around $\tau\approx 140$, which corresponds to a physical time of 17.5 minutes, approximately. The AW also presents a temporal evolution consistent with expectations in this case, with an amplification of the $b_y$ component up to $\tau\sim\tau_*$, where the $b_y$ component becomes primarily Alfv\'enic. During the wave-wave interaction around the DR, a small fraction of the energy transferred to the AW is returned back to the FMW, evidenced by the sudden drop in amplitude of $b_y$ just after $\tau =\tau_*$. Then, $b_y$ remains with an almost constant amplitude, slowly decrasing because of the weak ambipolar damping. Thus, in summary, the evolution exhibits some mixed wave behavior after and during the DR realizarion, but loses all FMW character shortly after surpassing the DR. This is expected, as the ambipolar damping time of the FMW is much shorter than that of the AW

On the other hand, the case with $K_y=10$ shows a completely different temporal evolution. To begin with, the larger value of $K_y$ makes the waves to have mixed properties from the beginning of the evolution. Both $b_x$ and $b_y$ display oscillations with a high frequency (the FMW frequency) overlaying a lower one (the AW frequency). Additionally, the effects of shear flow coupling are much more subtle than in the previous case, noticeable mainly through the change in the frequency of the $b_x$ component or the small change in amplitude in the $b_y$ component. These effects are completely absent from the $b_z$ component. However, the effects of ambipolar diffusion are  stronger than in the previous $K_y=1$ case. We observe that the FMW component of the oscillation, with a much higher frequency than in the previous case, becomes completely damped for a time around $\tau\sim100$, corresponding to 12.5 minutes in physical time, a shorter time scale than the obtained when $K_y=1$.  This reassures the conclusion that the larger the wavenumber, the faster the FMW is damped because of ambipolar effects. Later, for $\tau>100$, the $b_x$ perturbation  becomes a slowly damped wave of predominantly Alfv\'enic character, the $b_y$ perturbation also becomes an Alfv\'enic wave but with an almost constant amplitude, and the $b_z$ perturbation is completely damped as the FMW is fully dissipated by the ambipolar diffusion.

\begin{figure}
    \centering
    \includegraphics[width=0.9\hsize]{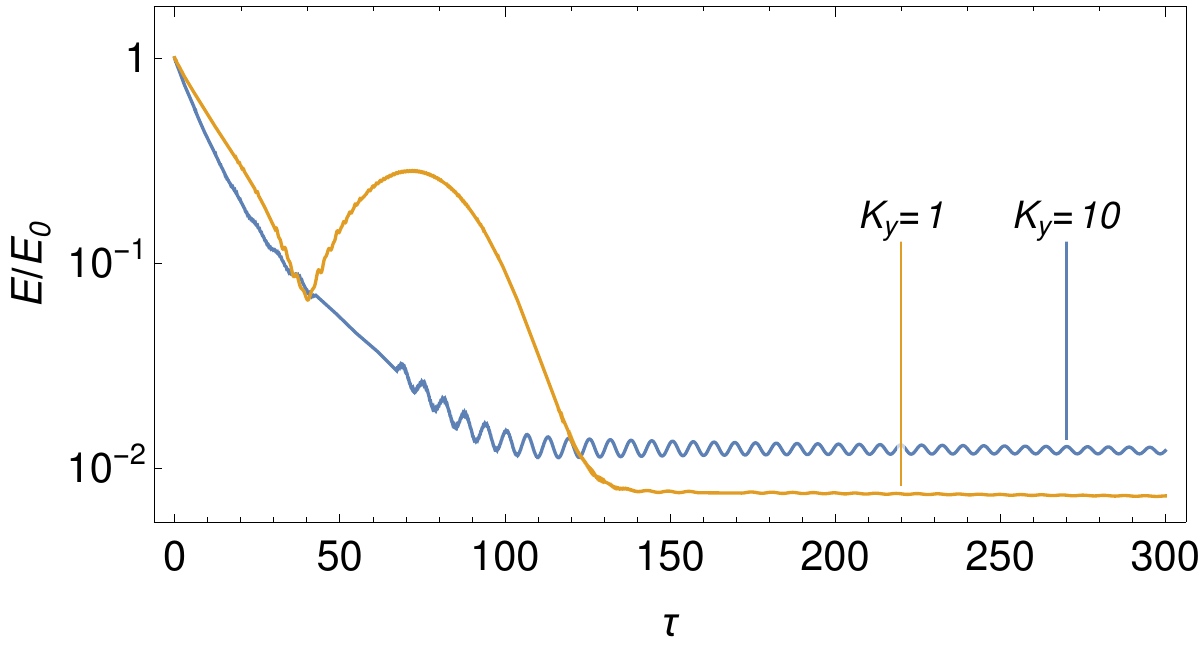}
    \caption{Temporal dependence of the normalized total energy, $E(\tau)/E(0)$, in the dissipative FMW-AW transformation for the solar prominence thread conditions of Section~\ref{app_threads} with $K_y=1$ (orange line) and $K_y=10$ (blue line).}
    \label{fig:Thread_case_Energy}
\end{figure}

The differences between both cases also become evident from the  energy temporal evolution displayed in Figure \ref{fig:Thread_case_Energy}. The first noticeable aspect is the absence of a local energy minimum around $\tau=\tau_*$ in the total energy curve of the $K_y=10$ case, as opposed to the case with $K_y=1$. Additionally, the dissipation of energy is faster in the case with larger $K_y$, despite maintaining a larger amount of residual energy associated with the longitudinal flow induced by ambipolar diffusion.

We conclude this application by summarizing some important points. Firstly, we found that the coupling and energy exchange between FMW and AW is possible in prominence threads and other similar structures, such as chromospheric spicules. Secondly, due to the ambipolar diffusion, the FMW have a much shorter lifespan than the AW, which persist for longer times. This is caused by the effect of the shear flow on the effective normal wavenumber, $K_x$. As $K_x$ increases (in absolute value) for large times, diffusion becomes more and more efficient for the FMW. Lastly, we observed that for large $K_y$ the shear-induced wave coupling loses relevance, but it is important for $K_y$ values consistent with kink modes.

\section{Concluding remarks}

We have shown that the energy exchanges and wave-wave interactions arising in a plasma with a shear flow are complex processes that are governed by multiple factors related to both the waves and the flow. The situation becomes even more complex when a dissipative effect, such as ambipolar diffusion, is included. For instance, we found that ambipolar diffusion introduces a new coupling mechanism. Although the derived governing equations are general, we have restricted ourselves to the FMW-AW transformation and have  conducted an exploratory investigation of how the shear flow and the ambipolar diffusion affects such transformation. The  model has been applied to a situation representative of a prominence thread, finding that the conditions that occur in these structures are suitable for the coupling and energy exchange between waves. In view of the obtained results, we plan to investigate further in this direction. As this coupling impacts on the evolution of the oscillations, it may be relevant to take its effects into account when interpreting the observations. 

It is important to note that the analysis carried out in this work is linear in nature. Consequently, nonlinear effects, such as the variation of the flow due to the energy exchange between the waves and the flow are absent. Besides, it should be recognized that considering these nonlinear effects may lead to substantial changes in the solutions as the flow itself evolves. Furthermore, while the nonmodal analysis is a valuable tool for obtaining solutions and describing the time evolution of the waves, it is essential to note that they are solutions in Fourier space. To obtain solutions in physical space, an inverse spatial Fourier transform is required. However, this is challenging due to the numerical nature of the results.

Several paths emerge to further extend the investigations conducted here. The immediate one is exploring the case with $\varepsilon\neq0$, as it introduces the SMW into the scene. This investigation is ongoing and will be presented in the continuation of this paper. The analysis of the role of the Hall term is also interesting. In another future work, we may use nonlinear numerical simulations to consider the full problem beyond the linear analysis done here. To this end, the nonlinear numerical simulations  of waves and flows in solar prominences by \citet{ofman2020,ofman2023} can be useful references.

\begin{acknowledgement}
This publication is part of the R+D+i project PID2023-147708NB-I00, funded by MCIN/AEI/10.13039/501100011033 and by FEDER, EU. This publication is part of the PREP2023-001251 grant, funded by MCIN/AEI/10.13039/501100011033 and the ESF+.  RS would like to thank Marc Carbonell for his help with Mathematica and for obtaining some of the initial results.
\end{acknowledgement}

\bibliographystyle{aa}
\bibliography{Bibliography}

\end{document}